\documentclass{aa}  
\usepackage{graphicx}
\bibpunct{(}{)}{;}{a}{}{,} % to follow the A&A style
\usepackage{txfonts}
\usepackage{lscape}
\usepackage{ulem}
\begin{document}

%-----------------------------------------------------------------------

\title{Dust properties in the cold and hot gas phases of the
  ATLAS$^{\rm 3D}$ early-type galaxies as revealed by AKARI}
% \subtitle{}

\author{T.\ Kokusho\inst{1} \and H.\ Kaneda\inst{1} \and M.\
  Bureau\inst{2,3} \and T.\ Suzuki\inst{1} \and K.\ Murata\inst{4} \and
  A.\ Kondo\inst{1} \and M.\ Yamagishi\inst{5} \and T.\ Tsuchikawa\inst{1}
  \and T.\ Furuta\inst{1}}

\institute{Graduate School of Science, Nagoya University, Chikusa-ku,
  Nagoya 464-8602, Japan\\
  \email{kokusho@u.phys.nagoya-u.ac.jp}
  \and
  Sub-department of Astrophysics, Department of Physics, University of
  Oxford, Denys Wilkinson Building, Keble Road, Oxford OX1 3RH, UK
  \and
  Yonsei Frontier Lab and Department of Astronomy, Yonsei University,
  50 Yonsei-ro, Seodaemon-gu, Seoul 03722, Republic of Korea
  \and
  Department of Physics, School of Science, Tokyo Institute of Technology,
  2-12-1 Ohokayama, Meguro, Tokyo, 152-8551, Japan
  \and
  Institute of Space and Astronautical Science, Japan Aerospace
  Exploration Agency, 3-1-1 Yoshinodai, Chuo-ku, Sagamihara, Kanagawa
  252-5210, Japan}

% \date{Received September 15, 1996; accepted March 16, 1997}

% \abstract{}{}{}{}{} 
\abstract
% context heading (optional)
{The properties of the dust in the cold and hot gas phases of
  early-type galaxies (ETGs) are key to understand ETG evolution.}
% aims heading (mandatory)
{We thus conducted a systematic study of the dust in a large sample of
  local ETGs, focusing on relations between the dust and the
  molecular, atomic, and X-ray gas of the galaxies, as well as their
  environment.}
% methods heading (mandatory)
{We estimated the dust temperatures and masses of the $260$ ETGs from
  the ATLAS$^{\rm 3D}$ survey, using fits to their spectral energy
  distributions primarily constructed from AKARI measurements. We also
  used literature measurements of the cold (CO and \ion{H}{i}) and
  X-ray gas phases.}
% results heading (mandatory)
{Our ETGs show no correlation between their dust and stellar masses,
  suggesting inefficient dust production by stars and/or dust
  destruction in X-ray gas.
  The global dust-to-gas mass ratios of ETGs
  are generally lower than those of late-type galaxies, likely due to
  dust-poor \ion{H}{i} envelopes in ETGs.
  They are also higher in Virgo Cluster ETGs than in group and field
  ETGs, but the same ratios measured in the central parts of the galaxies
  only are independent of galaxy environment.
  Slow-rotating ETGs have systematically lower dust masses than
  fast-rotating ETGs.
  The dust masses and X-ray luminosities
  are correlated in fast-rotating ETGs, whose star formation rates
  are also correlated with the X-ray luminosities. }
% conclusions heading (optional), leave it empty if necessary
{The correlation between dust and X-rays in fast-rotating ETGs appears
  to be caused by residual star formation, while slow-rotating ETGs
  are likely well evolved, and thus exhausting their dust.
  These results appear consistent with the
  postulated evolution of ETGs, whereby fast-rotating ETGs form by
  mergers of late-type galaxies and associated bulge growth, while
  slow-rotating ETGs form by (dry) mergers of fast-rotating
  ETGs. Central cold dense gas appears to be resilient against ram
  pressure stripping, suggesting that Virgo Cluster ETGs may not
  suffer strong related star formation suppression.}


\titlerunning{Dust properties of the ATLAS$^{\rm 3D}$ early-type
  galaxies with AKARI}
\authorrunning{Kokusho et al.}

\maketitle

%-----------------------------------------------------------------------

\section{Introduction}
\label{sec:intro}

Early-type galaxies (ETGs), consisting of elliptical and lenticular
galaxies, are thought to be in a late stage of galaxy evolution. They
are dominated by old stellar populations, that produce little dust,
and are often filled with X-ray-emitting plasma \citep[e.g.][]{for79},
that can destroy dust by sputtering on a timescale of
$10^6$--$10^7$~yr \citep{dra79}. ETGs thus provide severe conditions
for dust survival, but it has been shown that many ETGs nevertheless
do possess a cold interstellar medium (ISM; e.g.\
\citealt{kna85,war86,kna89,kna96}). The far-infrared (FIR) emission
from dust in ETGs was first observed by the Infrared Astronomical
Satellite (IRAS; \citealt{kna89}), that revealed that about half of
all ETGs observed contained cold dust. Because most of the IRAS
detections were close to its detection limit, \citet{bre98} carefully
evaluated the background contamination in the IRAS maps and found that
$12${\%} of the ETGs were detected in dust emission above the $98${\%}
confidence level. Following IRAS, more sensitive and detailed
observations were performed by the Infrared Space Observatory (ISO),
{\it Spitzer}, AKARI, and {\it Herschel}, that found pervasive dust in
ETGs, with a detection rate of ${\approx}50{\%}$
\citep[e.g.][]{tem04,tem07,kan11,smi12}.

Cold gas is also detected in many ETGs
\citep[e.g.][]{kna85,war86,com07}. \citet{you11} performed a census of
the molecular gas in the $260$ local ETGs of the ATLAS$^{\rm 3D}$
survey \citep{cap11}, revealing prevalent molecular gas and a
detection rate of $22{\%}$. Using AKARI, \citet{kok17} recently
measured the FIR emission from dust in the ATLAS$^{\rm 3D}$ ETGs,
confirming that the dust and molecular gas are well correlated in
CO-detected objects. This suggests that the dust and molecular gas in
ETGs are physically connected, just as in late-type galaxies
(LTGs). \citet{dav15} measured the dust, atomic, and molecular gas
masses of $17$ ETGs showing dust lanes in optical images, revealing
that most are gas-rich compared to LTGs and that the dust-to-gas mass
ratio varies widely from galaxy to galaxy. As metal enrichment is
likely to have proceeded well in ETGs, this indicates that most
dust-lane ETGs have probably acquired their gas through gas-rich
(minor) mergers. Indeed many authors have argued that ETGs amass their
cold ISM through external paths, such as galaxy mergers and gas
accretion from the intergalactic medium
\citep[e.g.][]{kna89,sar06,dav11,lag14}. Dust growth in interstellar
space is also postulated as an internal channel for dust production in
ETGs \citep{mar13_2,hir15}.

Galaxy environment is an important factor that may determine the
properties and origins of the cold ISM in ETGs. For example, the
\ion{H}{i} detection rate drops significantly for ETGs in the Virgo
Cluster, where \ion{H}{i} gas in galaxies appears to be stripped away
by the hot intracluster medium \citep[e.g.][]{dis07}. Within the
framework of the ATLAS$^{\rm 3D}$ project, \citet{dav11} demonstrated
that about half of the sample ETGs in the field have
kinematically-misaligned gas and stars, indicating that they likely
acquired their cold ISM through external paths. On the other hand,
ETGs in the Virgo Cluster almost always have kinematically-aligned gas
and stars, suggesting cold ISM of internal origin. A combination of
the two effects likely explains this field-cluster dichotomy. First,
it is often argued that when galaxies are virialised in a galaxy
cluster, the high velocity dispersion across cluster galaxies reduces
the galaxy merger rate \citep[e.g.][]{van99}.
Second, any minor merger is in any case likely to be dry, as small
galaxies should be ram pressure-stripped of their gas upon crossing
the cluster \citep{dav11}.

Whether dust in ETG interstellar space is replenished through internal
or external paths, dust should suffer heating and destruction by the
diffuse X-ray plasma prevalent in ETGs \citep[e.g.][]{for85}. This
diffuse X-ray plasma is generally attributed to stellar mass loss,
heated to the kinematic temperature determined by the galaxy stellar
velocity dispersion through collisions with the ambient gas
\citep[e.g.][]{can87}.  Gravitational heating by the galaxy potential
and supernova explosions also likely contribute to the heating
\citep[e.g.][]{mat86}. As the dust destruction rate should be
proportional to the density of the X-ray plasma, dust emission is
expected to be anti-correlated with the X-ray emission in
ETGs. \citet{tem07} and \citet{smi12} examined the relation between
dust and X-rays in ETGs, but found no correlation. Hence the nature of
the dust destruction process(es) in ETGs is still unclear, or dust
replenishment by internal and/or external sources must compensate the
dust destruction. A possible caveat of the above studies, however, is
that their X-ray measurements may include X-ray emission from point
sources (such as low-mass X-ray binaries and active galactic nuclei,
AGN; e.g.\ \citealt{bor11}) in addition to the diffuse plasma. A more
careful analysis of the X-ray emission is therefore needed to
precisely evaluate the state of the dust in the diffuse X-ray plasma
of ETGs.

As suggested above, the physical state of the dust in the cold and hot
gas phases of ETGs may imprint clues about the history of their cold
ISM, itself the fuel for residual star formation
\citep[e.g.][]{com07,dav14}.  Hence it is crucial to understand the
properties of the dust in these various gas phases to reveal the
evolution of ETGs. In this paper, we therefore perform a systematic
study of the dust in the ATLAS$^{\rm 3D}$ ETGs for which atomic
(\ion{H}{i}), molecular (CO), and ionised (X-ray) gas measurements are
available. We describe the data in Sect.~\ref{sec:sample} and present
the derived dust parameters in Sect.~\ref{sec:result}. We discuss the
dust properties of our ETGs by combining our results with literature
data in Sect.~\ref{sec:discussion}, and we summarise our results in
Sect.~\ref{sec:con}.

%-----------------------------------------------------------------------

\section{Sample and data}
\label{sec:sample}

\subsection{ATLAS$^{\rm 3D}$ survey}
\label{sec:atlas3d}

To understand the formation and evolution histories of ETGs, the
ATLAS$^{\rm 3D}$ survey \citep{cap11} performed a systematic study of
a volume-limited sample of $260$ nearby ($M_{\rm K}<-21.5$ and
$D<42$~Mpc) morphologically-selected ETGs. Molecular gas
\citep{com07,cro11,you11,ala13} and atomic gas \citep{ser12}
observations are available in addition to systematic optical
integral-field spectroscopy with the SAURON instrument
\citep{bac01}. Following the kinematic classification first introduced
by \citet{ems07} and \citet{cap07}, \citet{ems11} separated the
ATLAS$^{\rm 3D}$ sample galaxies into fast- ($86{\%}$) and
slow-rotating ETGs ($14{\%}$). This kinematic classification is less
affected by inclination and more robust than the standard
morphological classification into elliptical and lenticular galaxies,
and \citet{ems11} found that $66${\%} of the ellipticals in the
ATLAS$^{\rm 3D}$ sample are in fact fast-rotating. We therefore adopt
this kinematic classification in the present study.

\subsection{AKARI all-sky survey}
\label{sec:ir}

Using the AKARI all-sky maps (\citealt{mur07,doi15}; Ishihara et al.,
in prep.), \citet{kok17} systematically measured the $9$, $18$, $65$,
$90$, and $140$~$\mu$m band fluxes of the ATLAS$^{\rm 3D}$ ETGs. They
performed aperture photometry within a circular aperture of radius
$R_{\rm aper}=\sqrt{(2R_{\rm e})^2+(1.5D_{\rm PSF})^2}$, where
$R_{\rm e}$ and $D_{\rm PSF}$ are respectively the effective radius of
the galaxy in the optical $B$ band \citep{cap11} and the full width at
half maximum of the AKARI point spread function (PSF) in the relevant
band \citep{ish10,tak15}. Combining the AKARI measurements with
Wide-field Infrared Survey Explorer \citep[WISE;][]{cut13} and Two
Micron All Sky Survey (2MASS; \citealt{skr06}; \citealt{gri15}) data,
\citet{kok17} decomposed the resulting spectral energy distributions
(SEDs) into polycyclic aromatic hydrocarbon (PAH), warm dust, and cold
dust emission. For the dust emission, \citet{kok17} used a
two-temperature modified blackbody model with fixed emissivity
power-law index $\beta=2$. For the $68$ galaxies robustly detected
(i.e.\ with a signal-to-noise ratio $S/N>3$) in $2$ or $3$ of the
three AKARI FIR bands ($65$, $90$, and $140$~$\mu$m), the dust
temperatures were constrained by the fits, while for the remaining
galaxies robustly detected in only one or no FIR bands, the dust
temperatures were fixed to the means of the best-fit temperatures of
the $68$ galaxies detected in $2$ or $3$ FIR bands. In total, the SED
fits of $231$ galaxies are accepted at the $90{\%}$ confidence
level. We calculate their dust masses in Sect.~\ref{sec:dust}.

\subsection{X-ray data}
\label{sec:x-ray}

To evaluate the diffuse X-ray emission of ETGs, it is essential to
exclude contamination by X-ray point sources
\citep[e.g.][]{bor11}. Several studies have estimated the diffuse
X-ray emission of the ATLAS$^{\rm 3D}$ ETGs using $Chandra$ and its
unprecedented angular resolution \citep{sar13,kim15,su15,gou16}, thus
enabling them to mask out X-ray point sources and derive accurate
diffuse X-ray fluxes. Here, we adopt the X-ray measurements of
\citet{su15}, who measured the diffuse X-ray emission from the $42$
ATLAS$^{\rm 3D}$ ETGs observed with $Chandra$ for no less than
$15$~ks. They used a circular aperture of radius
$R_{\rm aper}=2R_{\rm e}$, identical to our infrared photometry
apertures except for our inclusion of the AKARI PSF extents.

%-----------------------------------------------------------------------

\begin{figure*}
  \centering
  \includegraphics[width=14cm]{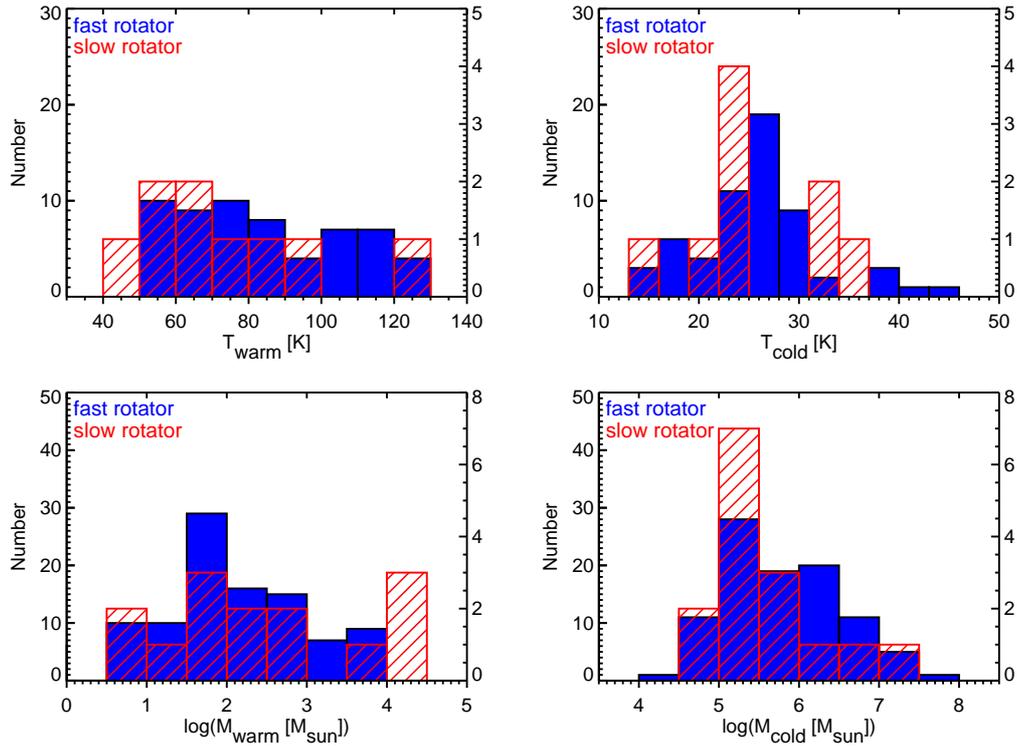}
  \caption{{\it Top}: histograms of the warm and cold dust
    temperatures for the $68$ ATLAS$^{\rm 3D}$ ETGs robustly detected
    in $2$ or $3$ of the three AKARI FIR bands. Blue (left vertical
    axes) and red (right vertical axes) represent fast- and
    slow-rotating galaxies, respectively. {\it Bottom}: as for the {\it top}
    panels, but for the warm and cold dust masses of the $111$
    ATLAS$^{\rm 3D}$ ETGs robustly detected in at least one AKARI FIR
    band.}
  \label{fig:hist}
\end{figure*}

\begin{figure*}
  \centering
  \includegraphics[width=8cm]{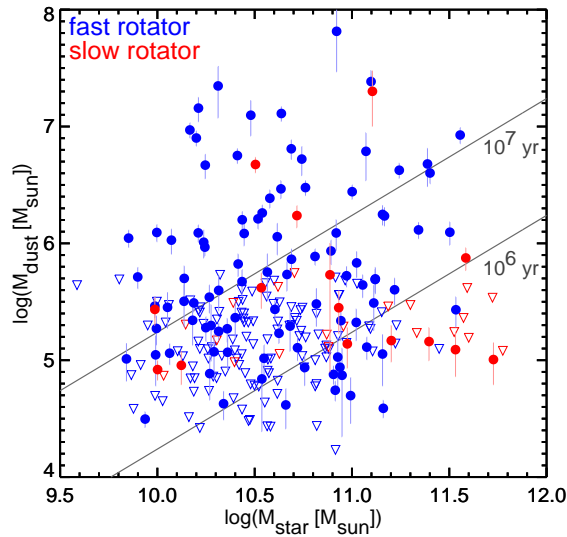}
  \caption{Total dust masses versus stellar masses. Solid circles
    indicate galaxies robustly detected in at least one AKARI FIR
    band. Downward open triangles indicate $3{\sigma}$ dust mass upper
    limits for the galaxies not robustly detected in any AKARI FIR
    band. Blue and red symbols represent fast- and slow-rotating
    galaxies, respectively. The gray lines show the dust masses
    expected from the balance between dust production from stellar
    mass loss and dust destruction from sputtering in an X-ray plasma,
    for the two dust destruction timescales indicated (see
    Sect.~\ref{sec:dust-star}).}
  \label{fig:dust-star}
\end{figure*}

\section{Results}
\label{sec:result}

\subsection{Far-infrared detection rates}
\label{sec:detect}

\citet{kok17} detected FIR dust emission in $45{\%}$ of the
ATLAS$^{\rm 3D}$ galaxies, where a FIR detection is defined as a
robust detection in at least one of the three AKARI FIR bands. The FIR
detection rate is $27{\%}$ and $52{\%}$ for the ATLAS$^{\rm 3D}$
elliptical and lenticular galaxies, respectively, showing that
lenticulars are more likely to contain cold dust. This is consistent
with previous studies, where the dust detection rate is $2$--$3$ times
higher in lenticulars than in ellipticals
\citep[e.g.][]{kna89,smi12,dis13}. On the other hand, the FIR
detection rate is $42{\%}$ and $46{\%}$ for the ATLAS$^{\rm 3D}$ fast-
and slow-rotating ETGs, respectively. The same trend is found from
$Herschel$ observations of local ETGs, where respectively $43{\%}$ and
$45{\%}$ of fast- and slow-rotating ETGs are robustly detected in the
$250$~$\mu$m band \citep{smi12}. However the number of fast- and
slow-rotating ETGs in these studies is very different, with
respectively $224$ versus $36$ in our sample and $51$ versus $11$ in
\citet{smi12}. The above trend should therefore be confirmed with a
larger number of slow-rotating ETGs.

\subsection{Dust temperatures and masses}
\label{sec:dust}

We estimate the dust masses of our sample ETGs using the
\citeauthor{kok17}'s (\citeyear{kok17}) SED fits. Their dust model is
described by
\begin{equation}
F_{\nu}=\frac{{\kappa}_{\nu}\,M_{\rm d}\,B{_\nu}(T_{\rm d})}{D^2}
\end{equation}
and
\begin{equation}
M_{\rm d}=\frac{F_{\nu}\,D^2}{{\kappa}_{\nu}\,B_{\nu}(T_{\rm d})}\,\,,
\end{equation}
where $F_{\nu}$ is the flux density in each band, ${\kappa}_{\nu}$ the
dust mass absorption coefficient parametrised as
${\kappa}_{\nu}\ {\propto}\ {\nu}^2$, $M_{\rm d}$ the dust mass,
$T_{\rm d}$ the dust temperature, $B_{\nu}(T_{\rm d})$ the Planck
function, and $D$ the distance to the galaxy. We adopt
${\kappa}_{140\ {\mu}{\rm m}}=13.9\ {\rm cm}^2\ {\rm g}^{-1}$
\citep{dra03} and the distances listed in \citet{cap11}. The derived
dust temperatures and masses are listed in Table.~A.1.

The top panels of Fig.~\ref{fig:hist} show the histograms of the warm
and cold dust temperatures ($T_{\rm warm}$ and $T_{\rm cold}$) for the
$68$ galaxies robustly detected in $2$ or $3$ of the three AKARI FIR
bands. There is a wide range of $T_{\rm warm}$ ($50$--$127$~K). Warm
dust emission in ETGs generally originates from circumstellar dust
around old stars \citep[e.g.][]{tem09}, but warm interstellar dust
heated by star formation activity and/or AGN has also been identified
\citep[e.g.][]{amb14}. The wide range of $T_{\rm warm}$ may thus
indicate that the dust heating mechanism varies from galaxy to
galaxy. A Kolmogorov-Smirnov (K-S) test indicates that for both
$T_{\rm warm}$ and $T_{\rm cold}$, the distributions of fast- and
slow-rotating ETGs are not statistically different. The mean value of
$T_{\rm cold}$ for the same $68$ ETGs is $26{\pm}1$~K higher than the
typical value for LTGs (${\lesssim}20$~K), a trend also reported by
\citet{smi12} and \citet{dis13}, who suggest that ETGs possess intense
stellar radiation fields due to their dense central populations of old
stars.

The bottom panels of Fig.~\ref{fig:hist} show the histograms of the
warm and cold dust masses ($M_{\rm warm}$ and $M_{\rm cold}$) for the
$111$ galaxies robustly detected in at least one AKARI FIR
band. $M_{\rm warm}$ and $M_{\rm cold}$ is estimated to be in the
range $10^{0.5}$--$10^{4.2}$~$M_\sun$ and
$10^{4.5}$--$10^{7.8}$~$M_\sun$, respectively, showing that the dust
mass in FIR-detected ETGs is dominated by the cold dust component,
likely found in the interstellar space of ETGs
\citep[e.g.][]{gou95,tem07,kok17}. The range of $M_{\rm cold}$ is in
agreement with that in previous studies. For example, \citet{smi12}
derived a $M_{\rm cold}$ range of $10^{5.0}$--$M^{7.1}$~$M_\sun$ for
local ETGs, while \citet{agi15} reported that their nearby ETGs have a
$M_{\rm cold}$ range of $10^{4.1}$--$10^{7.9}~M_\sun$. Again, the
differences between the $M_{\rm warm}$ and $M_{\rm cold}$
distributions of fast- versus slow-rotating ETGs are not statistically
significant.

\subsection{Dust and stellar masses}
\label{sec:dust-star}

Figure~\ref{fig:dust-star} shows the total dust masses
($M_{\rm dust}$), defined as the sums of $M_{\rm warm}$ and
$M_{\rm cold}$, plotted against the stellar masses ($M_\star$),
obtained from optical photometry and dynamically-measured
mass-to-light ratios in the ATLAS$^{\rm 3D}$ survey \citep{cap13}. The
galaxies not detected in any AKARI FIR band are shown as $3{\sigma}$
upper limits, where $\sigma$ is the uncertainty estimated from the SED
fit of each galaxy \citep{kok17}. The gray lines show $M_{\rm dust}$
predicted when the dust replenishment by stellar mass loss exactly
balances the destruction of dust grains by sputtering in an X-ray
plasma. To estimate these, we adopted a stellar mass loss rate
$\dot{M}=2.1\ {\times}\ 10^{-12}\ (L_K/L_{K,\odot})\ M_\odot\ {\rm
  yr}^{-1}$ \citep{kna92}, a typical $K$-band stellar mass-to-light ratio
$M_\star/L_K=1.21~M_\sun/L_{K,\sun}$ \citep{cap13}, and a dust
destruction timescale in an X-ray plasma of $10^6$ and $10^7$~yr,
respectively, for the two lines shown in Fig.~\ref{fig:dust-star},
the latter estimated by assuming a typical dust grain size of
$0.1$~$\mu$m and a gas density of $10^{-1}$ and $10^{-2}$~cm$^{-3}$,
respectively \citep{dra79}.

Figure~\ref{fig:dust-star} demonstrates that there is no correlation
between $M_{\rm dust}$ and $M_\star$ (linear correlation coefficient
$R=0.09$ and probability of deriving the observed $R$ if the null
hypothesis is true $p=0.35$ for FIR-detected ETGs), indicating that
stellar injection is not a significant source of dust in ETGs, as
suggested by previous studies \citep[e.g.][]{gou95,smi12}.
We also performed a correlation analysis including the $M_{\rm dust}$
upper limits using the generalised Kendall's $\tau$ test
implemented in the Astronomy SURVival Analysys (ASURV) package
\citep{iso86}, finding $p=0.09$. The above result therefore still holds
when including non-FIR-detected ETGs.
The figure
also shows that some ETGs are well above the expected $M_{\rm dust}$
shown by the gray lines, similarly suggesting that these galaxies
contain a large amount of dust that cannot be explained only by
stars. These results call for a dust supply from external sources
and/or efficient dust growth in the interstellar space of ETGs
\citep[e.g.][]{dav11,hir15}.

The mean value of $M_{\rm dust}/M_\star$ is $10^{-4.12{\pm}0.04}$ and
$10^{-4.56{\pm}0.09}$ for fast- and slow-rotating ETGs, respectively,
while it is ${\sim}10^{-3}$--$10^{-2}$ for LTGs \citep[e.g.][]{cor12},
systematically larger than for ETGs and suggesting that ETGs in
general are relatively poor in dust. \citet{smi12} showed that the
dust content of galaxies becomes smaller toward earlier types along
the Hubble sequence, but our results further suggest that
fast-rotating ETGs are richer in dust than slow-rotating ETGs. This
result is in agreement with the fact that fast-rotating ETGs tend to
possess relatively more massive cold ISM and associated ongoing star
formation \citep[e.g.][]{you11,dav14,kok17}.

%-----------------------------------------------------------------------

\section{Discussion}
\label{sec:discussion}

\subsection{Dust and cold gas}
\label{sec:dust-gas}

\begin{figure*}
  \centering
  \includegraphics[width=14cm]{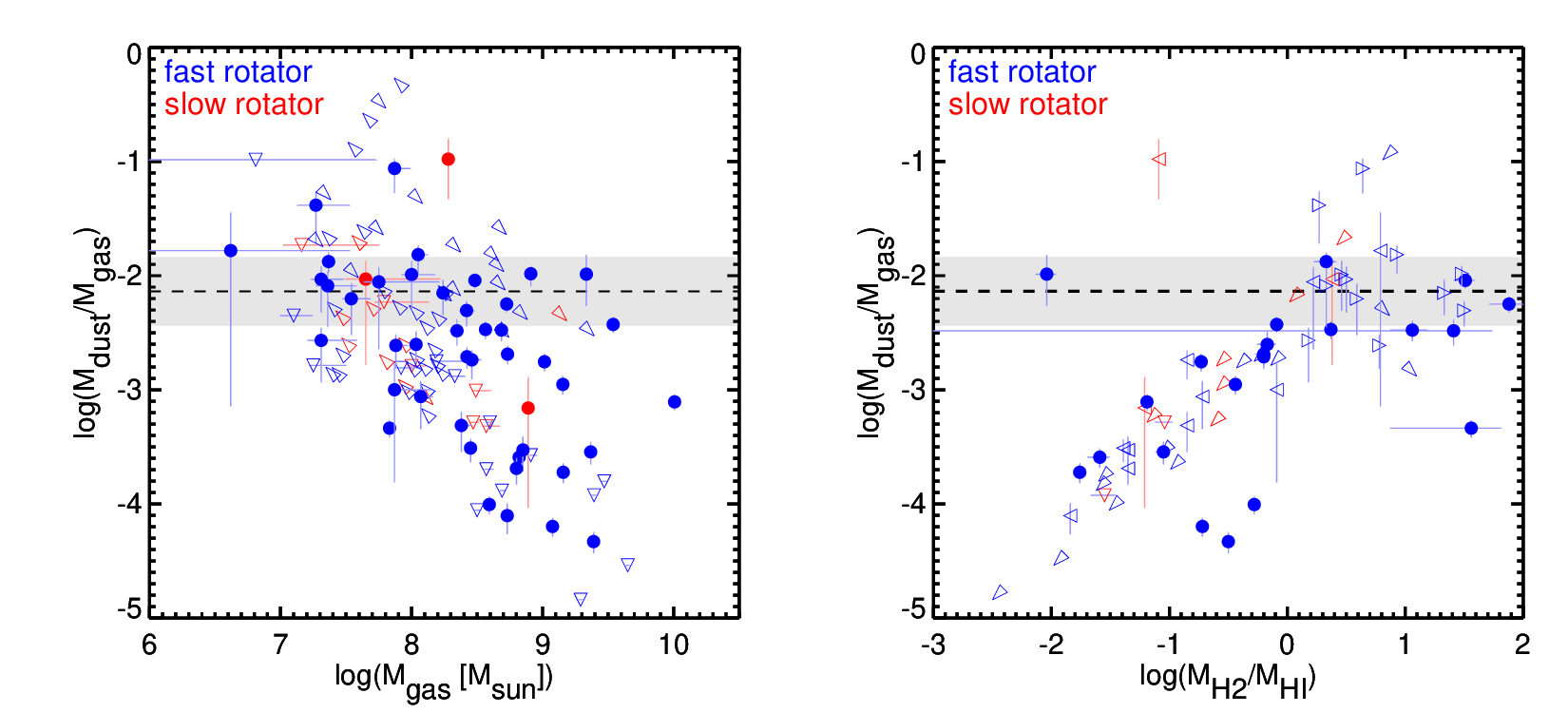}
  \caption{{\it Left}: global dust-to-gas mass ratios versus global
    cold gas masses ($M_{\rm gas}\equiv M_{{\rm
        H}_2}+M_\ion{H}{i}$). Solid circles show galaxies robustly detected
    in at least one AKARI FIR band and in at least one of H$_2$ and
    \ion{H}{i} (but observed in both), while downward open triangles show
    $3{\sigma}$ total dust mass upper limits for galaxies not robustly
    detected in any AKARI FIR band. Triangles pointing toward the
    upper-left show global cold gas upper limits (i.e.\ the sums of
    the $3{\sigma}$ upper limits on the global \ion{H}{i} and H$_2$
    masses) and are formally all global dust-to-gas mass ratio lower
    limits. Triangles pointing toward the bottom-right show H$_2$
    masses only and are formally all global dust-to-gas mass ratio
    upper limits and global cold gas mass lower limits. {\it Right}:
    global dust-to-gas mass ratios versus global H$_2$-to-\ion{H}{i}
    mass ratios. Solid circles show galaxies robustly detected in at least
    one AKARI FIR band and in both H$_2$ and \ion{H}{i}, while 
    $3\sigma$ H$_2$ (resp.\ \ion{H}{i}) upper limits are shown as
    leftward (resp.\ rightward) open triangles. Triangles pointing toward the
    bottom-left (resp.\ bottom-right) show $3\sigma$ H$_2$ (resp.\
    \ion{H}{i}) upper limits for galaxies not robustly detected in any
    AKARI FIR band and are formally all global dust-to-gas mass ratio
    upper limits. In both panels, blue and red symbols represent
    respectively fast- and slow-rotating galaxies, and the black
    dotted line and gray shaded region show respectively the relation
    of our Galaxy \citep{dra07} and a factor $2$ spread around it.}
  \label{fig:dust-gas}
\end{figure*}

\begin{figure*}
  \centering
  \includegraphics[width=14cm]{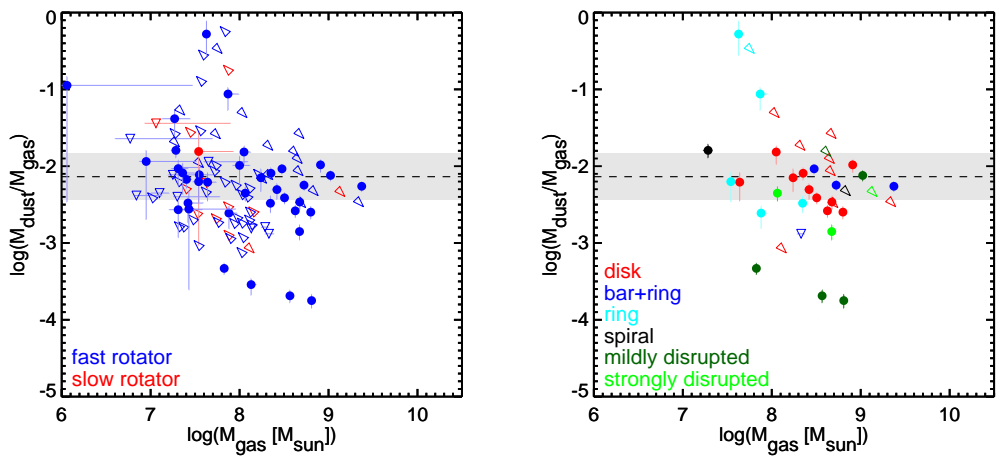}
  \caption{As for the {\it left} panel of Fig.~\ref{fig:dust-gas}, but for
    the central dust-to-gas mass ratios versus central cold gas
    masses, where the \ion{H}{i} measurements are from the central
    regions of the galaxies only
    ($\approx35^{\arcsec}\times45^{\arcsec}$; \citealt{you14}).  Data
    points are colour-coded according to the galaxies' specific stellar
    angular momenta (fast- and slow-rotators; {\it left}) and CO
    morphologies (\citealt{ala13}; {\it right}).}
  \label{fig:dust-gas2}
\end{figure*}

\begin{figure*}
  \centering
  \includegraphics[width=14cm]{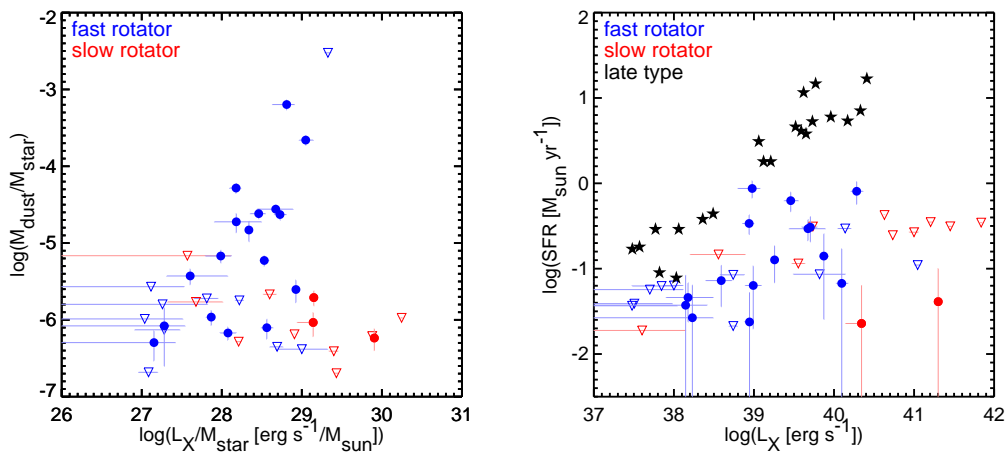}
  \caption{{\it Left}: total dust masses versus diffuse X-ray plasma
    luminosities, both normalised by the stellar masses, for the
    galaxies observed with $Chandra$ by \citet{su15}. {\it Right}:
    star formation rates versus diffuse X-ray plasma luminosities. The
    symbols are as in Fig.~\ref{fig:dust-star}, but the black stars
    in the {\it right} panel show robustly detected late-type galaxies
    \citep{min12}.}
  \label{fig:dust-x}
\end{figure*}

\begin{figure*}
  \centering
  \includegraphics[width=14cm]{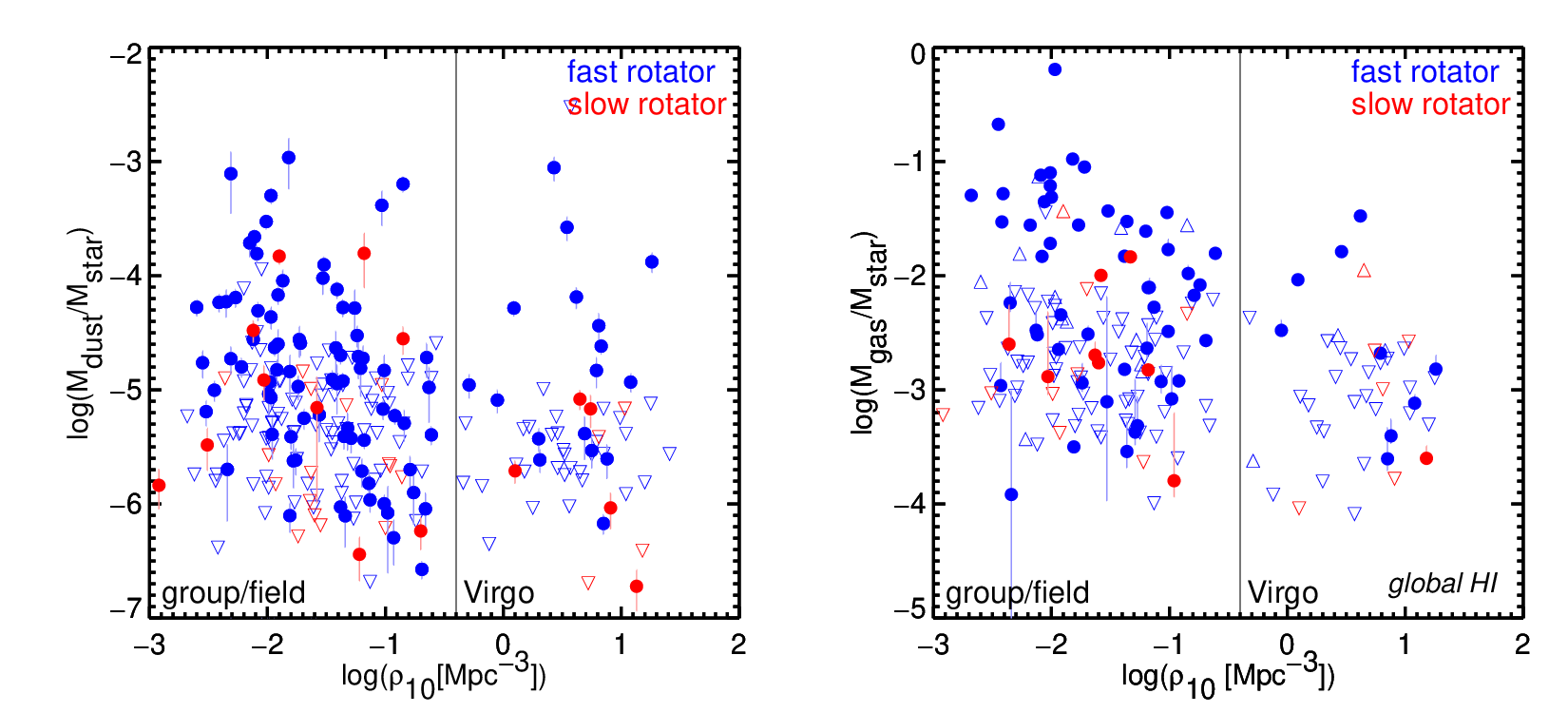}
  \includegraphics[width=14cm]{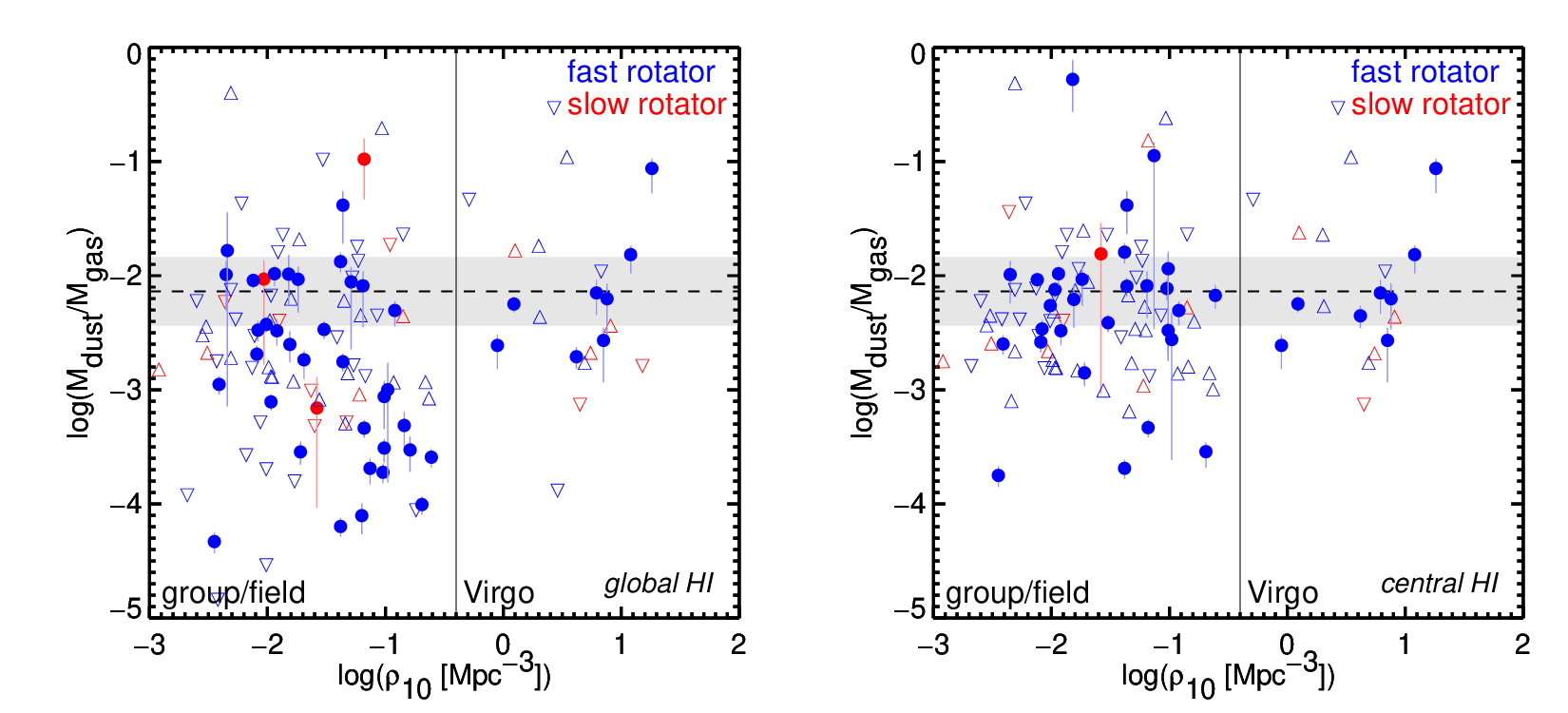}
  \includegraphics[width=14cm]{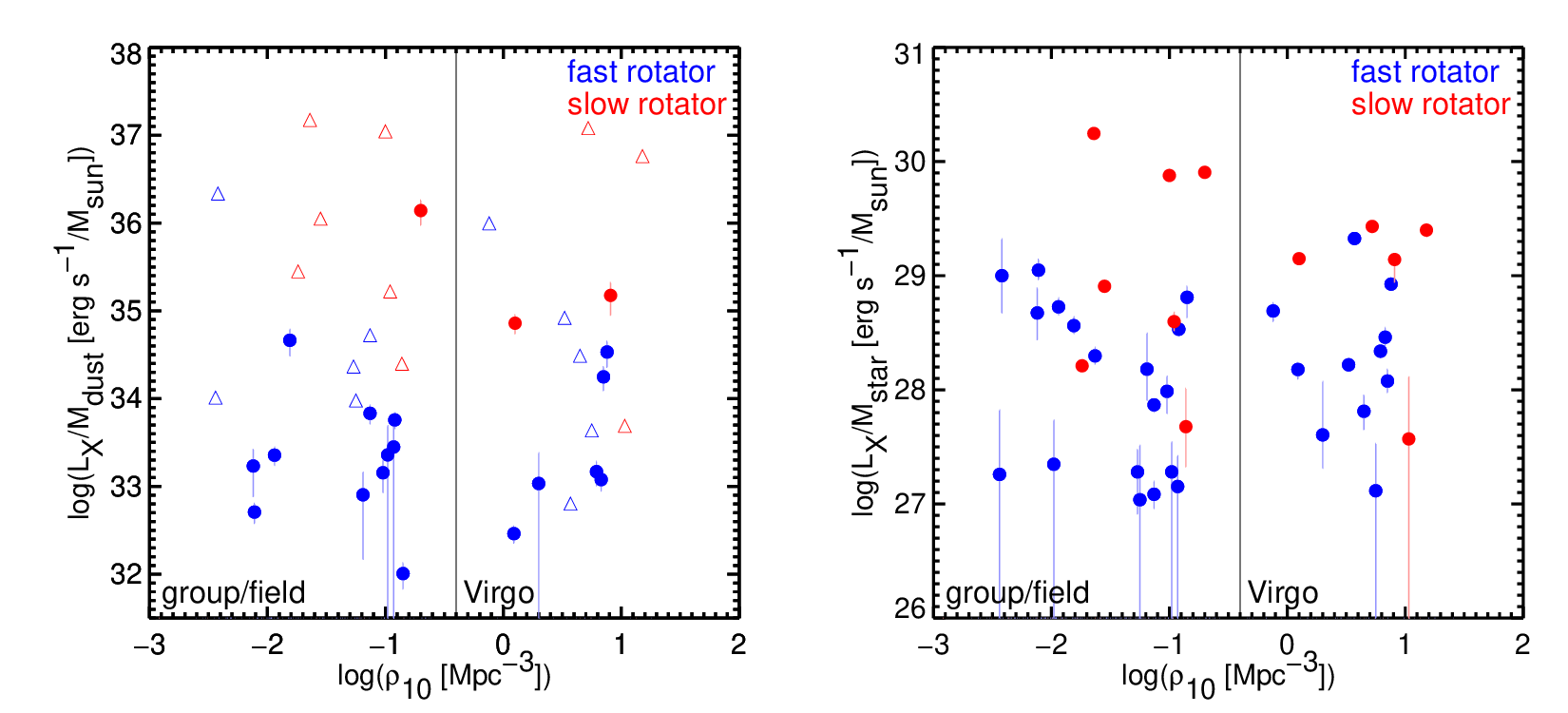}
  \caption{{\it Top-left}: total dust-to-stellar mass ratios versus
    local galaxy densities ${\rho}_{10}$ \citep{cap11b}. The symbols
    are as in Fig.~\ref{fig:dust-star}. {\it Top-right}: as the {\it
      top-left} panel, but for the global gas-to-stellar mass
    ratios. Solid circles show galaxies detected in at least one of H$_2$
    and \ion{H}{i} (but observed in both). Downward open triangles show
    global cold gas mass upper limits (i.e.\ the sums of the
    $3{\sigma}$ upper limits on the global \ion{H}{i} and H$_2$
    masses). Upward triangles show H$_2$ masses only and are formally
    all global gas-to-stellar mass ratio lower limits. {\it
      Middle-left}: global dust-to-gas mass ratios versus
    ${\rho}_{10}$. Solid circles show galaxies robustly detected in at least
    one AKARI FIR band and in at least one of H$_2$ and \ion{H}{i}
    (but observed in both). Downward open triangles show $3{\sigma}$ total
    dust mass upper limits for galaxies not robustly detected in any
    AKARI FIR band. Galaxies observed and detected in H$_2$ only are
    also shown as downward triangles and are formally all global
    dust-to-gas mass ratio upper limits and global cold gas mass lower
    limits. Upward triangles show global cold gas mass upper limits
    and are formally all global dust-to-gas mass ratio lower limits.
    {\it Middle-right}: as the {\it middle-left} panel, but for the
    central dust-to-gas mass ratios. {\it Bottom-left}: X-ray
    luminosity-to-dust mass ratios versus ${\rho}_{10}$. Solid circles show
    galaxies robustly detected in at least one AKARI FIR band. Upward
    open triangles show $3{\sigma}$ total dust mass upper limits for
    galaxies not robustly detected in any AKARI FIR band and are
    formally all X-ray luminosity-to-dust mass ratio lower
    limits. {\it Bottom-right}: as the {\it bottom-left} panel, but
    for the X-ray luminosity-to-stellar mass ratios. In all panels,
    the solid line at $\log(\rho_{10}/{\rm Mpc}^3)=-0.4$ divides Virgo
    Cluster galaxies from group and field galaxies. The horizontal lines and
    gray shaded regions in the {\it middle} panels are as in
    Fig.~\ref{fig:dust-gas}.}
  \label{fig:density}
\end{figure*}

\begin{figure*}
  \centering
  \includegraphics[width=14cm]{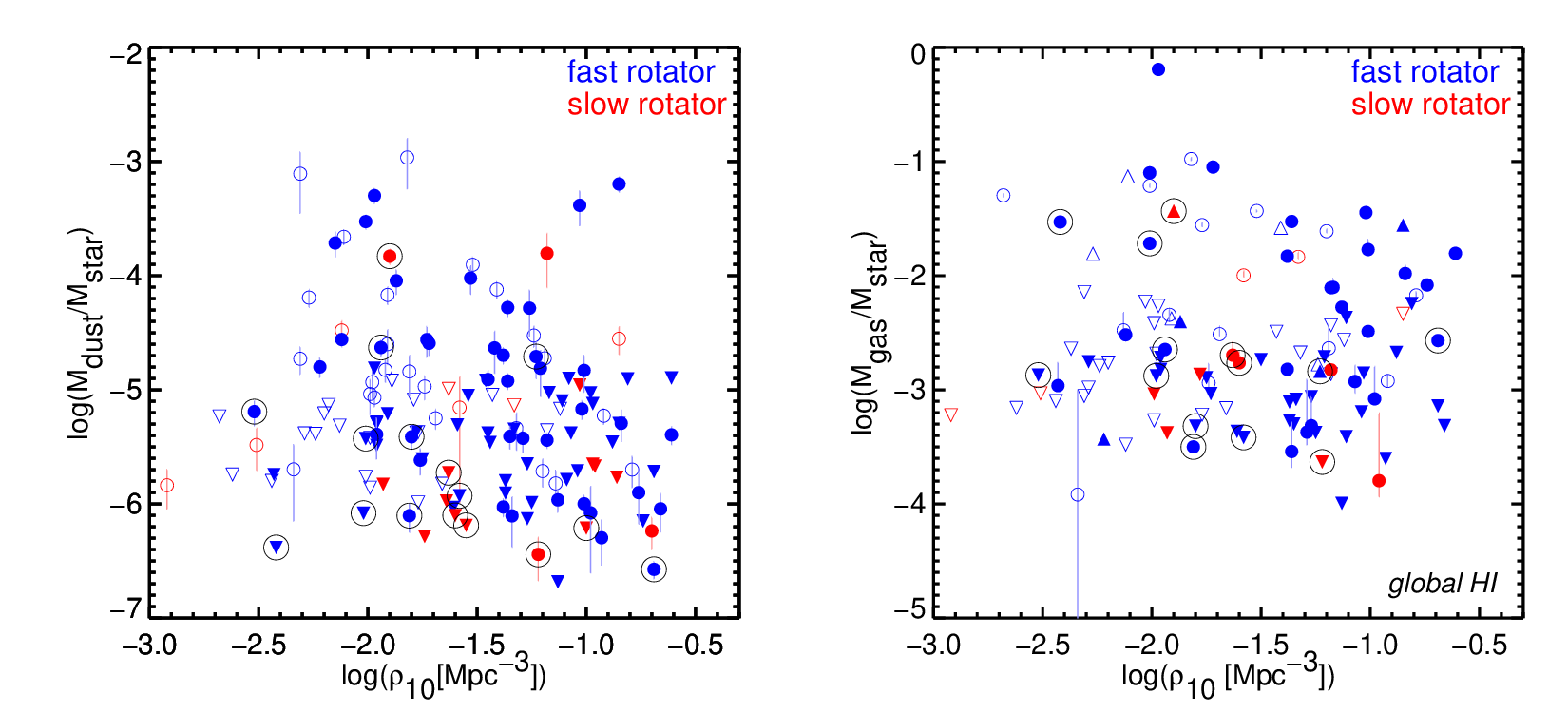}
  \includegraphics[width=14cm]{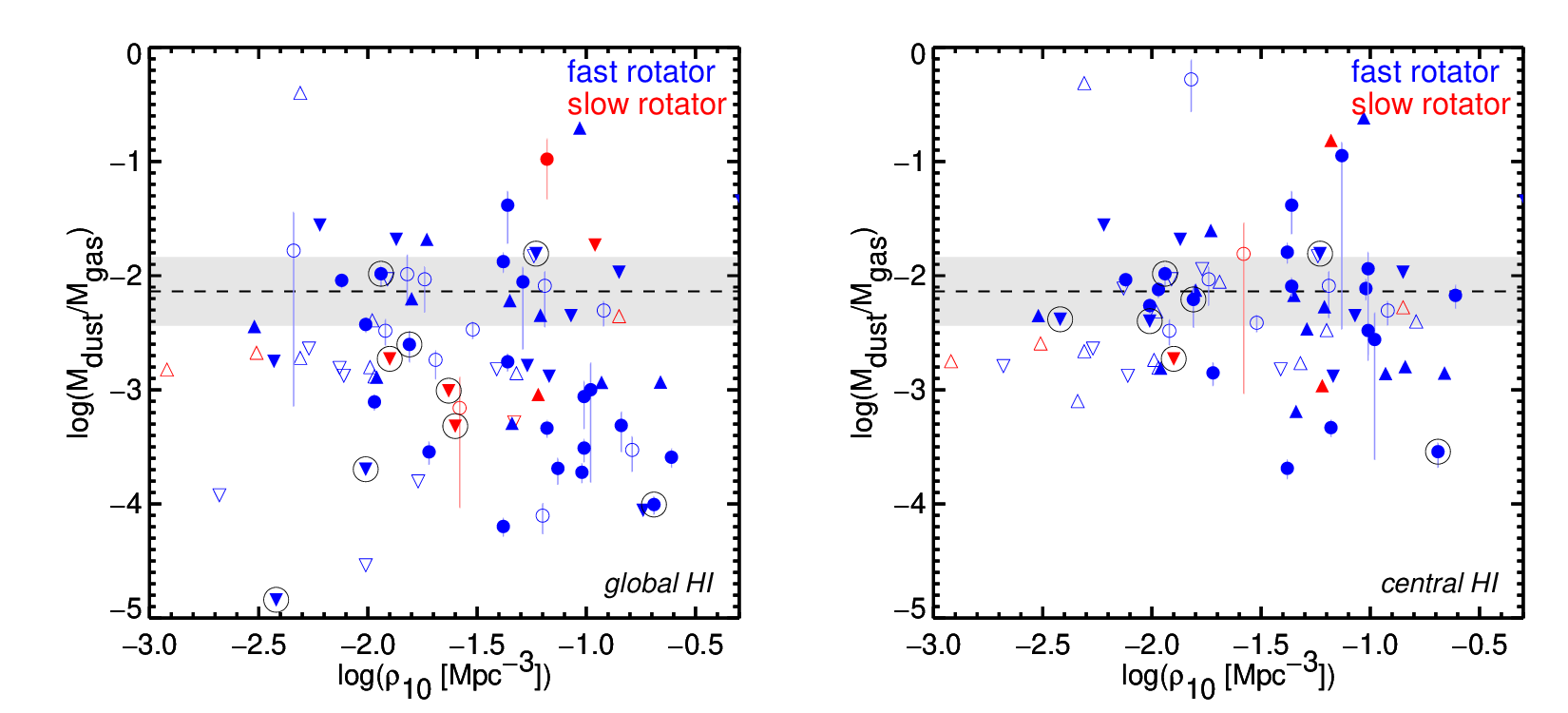}
  \includegraphics[width=14cm]{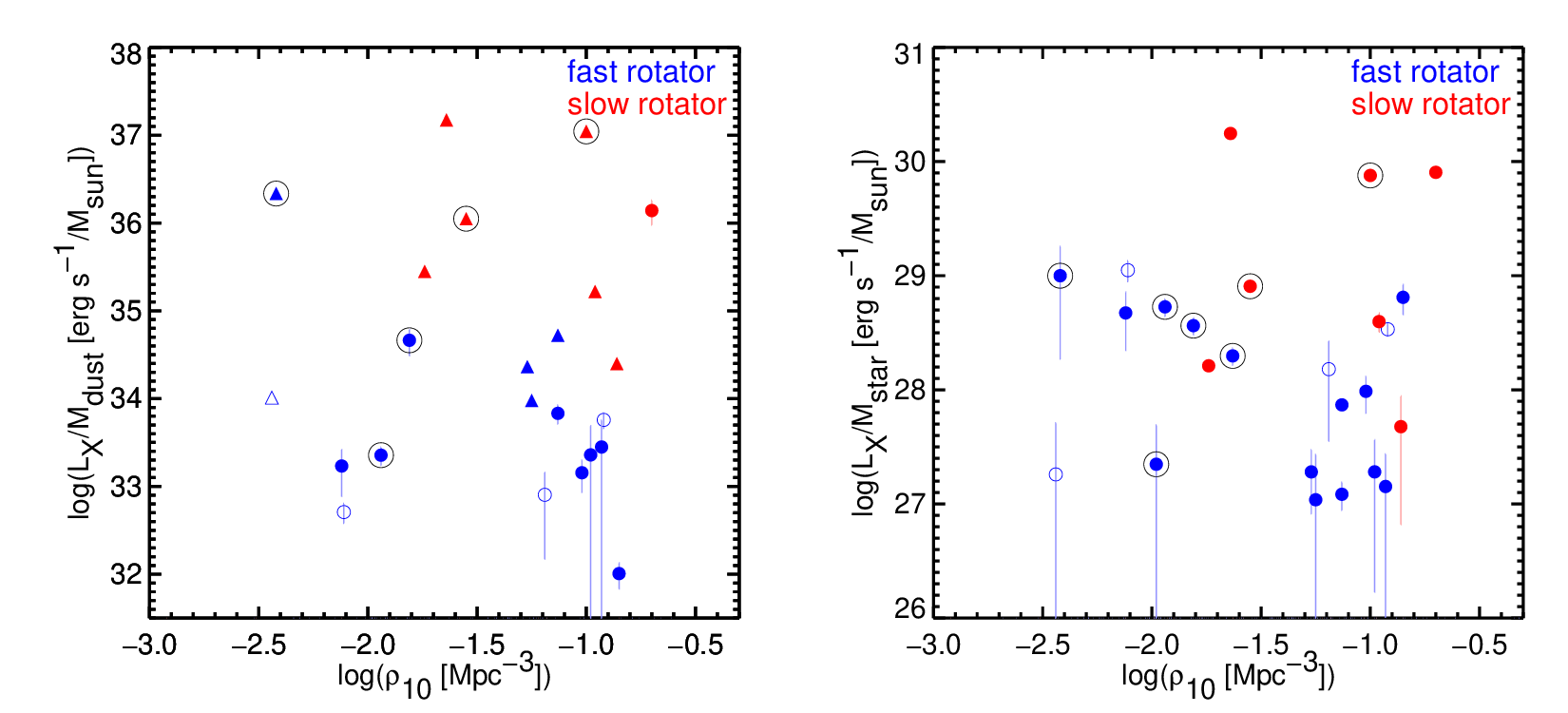}
  \caption{As for Fig.~\ref{fig:density}, but for galaxies at
    $\log(\rho_{10}/{\rm Mpc}^3)<-0.4$ only. Solid symbols show
    galaxies belonging to a group \citep{gar93}, while open symbols
    show those in the field. The optically-brightest galaxies
    identified by \citet{gar93} are indicated with additional larger
    open circles.}
  \label{fig:group}
\end{figure*}

The left panel of Fig.~\ref{fig:dust-gas} shows the global
dust-to-gas mass ratios ($M_{\rm dust}/M_{\rm gas}$) plotted
against the global cold gas masses ($M_{\rm gas}$), defined as the
sums of the molecular gas masses ($M_{{\rm H}_2}$) and the global
neutral hydrogen gas masses ($M_\ion{H}{i}$), where $M_{{\rm H}_2}$
was estimated for all ATLAS$^{\rm 3D}$ ETGs by \citet{you11}, who used
a Galactic CO-to-H$_2$ conversion factor \citep{dic86,str88,str04},
and $M_\ion{H}{i}$ was estimated for $166$ ATLAS$^{\rm 3D}$ ETGs at a
declination above $10^{\circ}$ by \citeauthor{ser12}
(\citeyear{ser12}; except for four objects near the Virgo Cluster
centre). The black dotted line shows the relation of our Galaxy
($M_{\rm dust}/M_{\rm gas}=0.0073$; \citealt{dra07}) and the gray
shaded region a factor $2$ spread around it, where most LTGs are found
\citep{dra07}. This figure clearly shows that while the scatter is
significant, the global dust-to-gas mass ratios of the
ATLAS$^{\rm 3D}$ ETGs are generally lower than those of LTGs,
suggesting that the cold dense ISM of ETGs is generally poor in dust.

The right panel of Fig.~\ref{fig:dust-gas} shows the relation between
the global $M_{\rm dust}/M_{\rm gas}$ and the global
$M_{{\rm H}_2}/M_\ion{H}{i}$ for the ATLAS$^{\rm 3D}$ ETGs detected in
at least one of H$_2$ and \ion{H}{i}. This clearly shows that galaxies
with low global $M_{\rm H_2}/M_\ion{H}{i}$ also have low global
$M_{\rm dust}/M_{\rm gas}$. However, molecular gas is generally
concentrated in the central regions of ETGs \citep[e.g.][]{ala13},
while \ion{H}{i} gas tends to be extended beyond the stellar body
\citep[e.g.][]{oos10}, so one possible explanation of the trend
observed is that an appreciable amount of dust is present beyond the
infrared photometric aperture we used. To test this possibility, we
compared our AKARI aperture sizes with the \ion{H}{i} extents of the
$24$ ATLAS$^{\rm 3D}$ ETGs whose \ion{H}{i} morphology is defined as
{\it large disk} in \citet{ser12}, finding that the \ion{H}{i} gas is
more extended than our largest AKARI aperture in $23$ out of these
$24$ ETGs. For each of these galaxies, we thus enlarged our aperture
to cover the entire \ion{H}{i} extent, and re-measured the FIR flux in
the $90$~$\mu$m band, the AKARI FIR band most sensitive to cold dust
emission \citep{kok17}. Re-estimating $M_{\rm dust}$ with these new
$90$~$\mu$m fluxes, the trend in the right panel of
Fig.~\ref{fig:dust-gas} does not change, i.e.\ galaxies with low
global $M_{\rm H_2}/M_\ion{H}{i}$ still have low global
$M_{\rm dust}/M_{\rm gas}$. This indicates that the \ion{H}{i}
envelopes of ETGs are likely to be intrinsically poor in dust. Such
dust-deficient \ion{H}{i} envelopes are found in LTGs
\citep[e.g.][]{fer98}, but our data suggest that they are also likely
to be present in ETGs.

Of course, it is known that dust tightly correlates with molecular gas
(the median value of $M_{\rm dust}/M_{\rm H_2}$ is $0.0075$ for
  FIR- and CO-detected ETGs;
see, e.g., \citealt{kok17} for the ATLAS$^{\rm 3D}$ ETGs), indicating
that dust and molecular gas are likely to coexist. The \ion{H}{i}
envelopes of ETGs may therefore be poor not only in dust, but also in
molecular gas (i.e.\ ETG dust is related to H$_2$ only and not
\ion{H}{i}).
Conversely, when dust is closely associated with dense molecular
clouds, the molecular gas may shield the dust from the hot X-ray
plasma and thus prevent its destruction via sputtering
\citep[e.g.][]{dej90}. This could explain why many ETGs still
possess a lot of dust (see Sect.~\ref{sec:dust-star}).

The left panel of Fig.~\ref{fig:dust-gas2} shows the relation
between the central (as opposed to global) $M_{\rm dust}/M_{\rm gas}$
and the central $M_{\rm gas}$, where we adopted \citeauthor{you14}'s
(\citeyear{you14}) $M_\ion{H}{i}$ measurements from the central
regions of the galaxies only
(${\approx }35^{\arcsec}{\times}45^{\arcsec}$) rather than total
measurements. This figure reveals that most of our ETGs have a central
$M_{\rm dust}/M_{\rm gas}$ similar to those of LTGs, suggesting
that their cold dense ISM are also similar to those of LTGs. As
metallicity is an important factor governing star formation (via gas
cooling), this result can naturally explain why ETGs have star
formation efficiencies similar to those of LTGs \citep{sha10,kok17}.

Nevertheless, the left panel of Fig.~\ref{fig:dust-gas2} also shows a
few ETGs with central $M_{\rm dust}/M_{\rm gas}$ significantly higher
or lower than those of LTGs. \citet{dav15} argued that external gas
accretion may cause the large $M_{\rm dust}/M_{\rm gas}$ scatter, as
ETGs are expected to generally have relatively high
$M_{\rm dust}/M_{\rm gas}$ due to secular metal enrichment. The gas
accretion histories of the ATLAS$^{\rm 3D}$ ETGs were studied by
\citet{ala13} through their CO distributions. They classified the CO
morphologies of the $40$ galaxies with interferometric (i.e.\
spatially-resolved) data into $6$ categories: mildly disrupted,
strongly disrupted, disk, bar+ring, ring, and spiral, suggesting that
galaxies with disrupted CO probably acquired their molecular gas
externally. We thus colour-coded our ETGs by their CO morphology in
the right panel of Fig.~\ref{fig:dust-gas2}. Interestingly, this
figure shows that the $3$ galaxies with the lowest central
$M_{\rm dust}/M_{\rm gas}$ all have mildly disrupted CO, suggesting
that more pristine external gas (e.g.\ from a lower mass accreted
galaxy) may have diluted their $M_{\rm dust}/M_{\rm gas}$. Conversely,
the $3$ galaxies with the highest central $M_{\rm dust}/M_{\rm gas}$
all have a CO ring (and the next $2$ a CO disk), known to form
naturally through secular processes (e.g.\ bar-driven;
\citealt{but96}) with no necessity of external gas accretion. Metal
enrichment could thus have proceeded naturally in those galaxies with
a CO ring and/or disk. This scenario should however be verified by
more sensitive, detailed, and numerous spatially-resolved
CO observations of ETGs.

\subsection{Dust and diffuse X-ray plasma}
\label{sec:dust-x}

The left panel of Fig.~\ref{fig:dust-x} shows the total dust masses
($M_{\rm dust}$) plotted against the diffuse X-ray plasma total
luminosities ($L_{\rm X}$) for the $42$ ATLAS$^{\rm 3D}$ ETGs observed
with {\it Chandra} by \citet{su15}. We normalised both quantities by
$M_\star$ as the diffuse X-ray plasma is supposed to be sustained by
the galaxies' gravitational potentials (and thus $L_{\rm X}$ should
scale with $M_\star$; e.g.\ \citealt{for85}).
Slow-rotating ETGs tend to show lower $M_{\rm dust}/M_\star$ than
fast-rotating ETGs, especially at higher $L_{\rm X}/M_\star$.
In contrast, fast-rotating ETGs show a clear correlation between
$M_{\rm dust}/M_\star$ and $L_{\rm X}/M_\star$ ($R=0.63$ and
$p<0.01$ for FIR-detected fast-rotating ETGs).
We again performed a correlation analysis including the
$M_{\rm dust}/M_\star$ upper limits using the generalised Kendall's
$\tau$ test \citep{iso86}, finding $p<0.01$. The above result therefore
still holds when including non-FIR-detected fast-rotating ETGs.
No correlation was found between $M_{\rm dust}$ and
$L_{\rm X}$ in the previous work of \citet{smi12} on nearby ETGs,
where they normalised both quantities by the optical $B$-band
luminosities and used $L_{\rm X}$ measured either from R\"{o}ntgen
Satellite (ROSAT; \citealt{osu01}) or $Chandra$ (nuclei;
\citealt{pel10}). However, $B$ band does not trace stellar masses as
well as $K$ band, and their X-ray luminosities are possibly
contaminated by low-mass X-ray binaries and/or AGN, which may hide a
relation between $M_{\rm dust}$ and $L_{\rm X}$ and explain the
negative result.

The right panel of Fig.~\ref{fig:dust-x} shows the relation between
star formation rate (SFR) and $L_{\rm X}$ for ATLAS$^{\rm 3D}$ ETGs,
where the SFRs were derived from PAH emission by \citet{kok17}. Also
shown are the $21$ LTGs for which \citet{min12} measured the diffuse
X-ray plasma emission in the same manner as \citet{su15} for the
ATLAS$^{\rm 3D}$ ETGs. The LTG SFRs were derived from a combination of
ultraviolet (UV) and FIR emission \citep{min12}.
The figure shows that our sample ETGs have systematically lower
SFRs than LTGs, likely due to their smaller cold gas fractions,
while they have star formation efficiencies similar to those of LTGs
\citep[see Sect.~\ref{sec:dust-gas}; e.g.][]{sha10,kok17}.
The figure shows a
tight correlation between the SFRs and $L_{\rm X}$ in LTGs, where
supernova remnants are likely the dominant X-ray source
\citep{min12}. However, fast-rotating ETGs also show a correlation
($R=0.58$ and $p=0.02$ for FIR-detected fast-rotating ETGs),
suggesting that their star formation activity
also contributes to the X-rays.
We again performed a correlation analysis including the SFR upper
limits using the generalised Kendall's $\tau$ test, finding
$p<0.01$. The above result therefore still holds when including
non-FIR-detected fast-rotating ETGs.
Assuming that $M_{\rm dust}$ traces
the cold dense ISM, $M_{\rm dust}/M_\star$ may then be indirectly
related to $L_{\rm X}/M_\star$ via residual star formation activity.

ETGs are thought to have a bottom-heavy stellar initial mass function
\citep[e.g.][]{dok10,cap12,kok17}, so that supernova explosions
perhaps do not dominate their X-ray emission. Another possibility is
young stellar objects (YSOs), that emit X-rays through magnetic
reconnection in their stellar magnetospheres and circumstellar disks
\citep[e.g.][]{fei99}. Using a median $L_{\rm X}=10^{29.4}$~erg
s$^{-1}$ from the Chamaeleon~I cloud YSOs \citep{fei93}, and the area
of the star-forming regions determined by \citet{dav14}, we can
estimate the number density of YSOs in the $9$ ETGs that have both
X-ray and CO interferometric observations. Assuming spherical
star-forming regions, the median value of the YSO number densities is
estimated to be $\approx3$~pc$^{-3}$, comparable to that of
star-forming clouds in our Galaxy ($\approx1$~pc$^{-3}$;
\citealt{hei10}). Hence the $M_{\rm
  dust}/M_\star$\,--\,$L_{\rm X}/M_\star$ correlation in fast-rotating
ETGs can be produced by their current star formation activity, whether
the X-rays originate from supernova explosions and/or YSOs.

Our results support the current preferred evolutionnary scenario for
ETGs \citep[e.g.][]{cap13b,pen17}, whereby fast-rotating ETGs are
thought to have evolved from LTGs through major or minor mergers that
trigger intense star formation and associated bulge growth. Star
formation is still ongoing in some of these galaxies, producing the
$M_{\rm dust}/M_\star$\,--\,$L_{\rm X}/M_\star$ correlation shown in
Fig.~\ref{fig:dust-x}. The diffuse X-ray plasma may have just
started to grow through secular (gas heating by stellar motions and
the gravitational potentials) or merger-induced \citep[e.g.][]{mon06}
processes. Slow-rotating ETGs are thought to have evolved from
fast-rotating ETGs via several (dry) major mergers, and to have grown
their diffuse X-ray plasma for a long time (thus destroying their cold
ISM). 
This scenario should however be verified by more sensitive
and numerous observations of dust and X-rays in slow-rotating ETGs.

\subsection{Galaxy environment and the cold ISM}
\label{sec:env}

The intracluster media of galaxy clusters are thought to strip the
cold ISM of member galaxies through ram pressure and to prevent
external cold gas accretion. In addition, when galaxies are virialised
within clusters, strong galaxy interactions are less likely due to the
high velocity dispersion across galaxies \citep[e.g.][]{van99}.
Galaxy groups are an environment intermediate between field and
  clusters, and most galaxy groups are known to possess an X-ray
  emitting intra-group medium \citep[IGM; e.g.][]{osu17}, that can
  heat and/or destroy the cold ISM of galaxies. In addition, as the
  velocity dispersion across galaxies in groups is smaller than that
  in clusters, galaxy mergers are more likely to occur in groups than
  in clusters \citep[e.g.][]{alo12}.
To investigate such environmental effects on the cold ISM of ETGs,
the top panels of Fig.~\ref{fig:density} show the $M_{\rm dust}/M_\star$
and global $M_{\rm gas}/M_\star$ plotted against the local volume
densities of our sample galaxies, measured in a sphere which is
centered on each galaxy and contains the $10$ nearest galaxy
neighbours (both ETGs and LTGs; $\rho_{10}$, \citealt{cap11b}). The
solid line ($\log(\rho_{10}/{\rm Mpc}^3)=-0.4$) divides Virgo Cluster
ETGs from group and field ETGs.
Although there is much scatter, the figures
show that $M_{\rm dust}/M_\star$ is independent of galaxy environment
while the global $M_{\rm gas}/M_\star$ tends to be lower in Virgo Cluster
ETGs than in group and field ETGs, supporting a scenario whereby the diffuse
\ion{H}{i} envelopes are more easily ram pressure-stripped than the
dust \citep[e.g.][]{cor12}.

The middle panels of Fig.~\ref{fig:density} show the global and
central $M_{\rm dust}/M_{\rm gas}$ plotted against $\rho_{10}$,
showing that the global $M_{\rm dust}/M_{\rm gas}$ is higher in
Virgo Cluster ETGs than in group and field ETGs while the central
$M_{\rm dust}/M_{\rm gas}$ is independent of galaxy environment.
To better quantify this trend, we performed a K-S test on the
global and central $M_{\rm dust}/M_{\rm gas}$, comparing Virgo
Cluster ETGs and group and field ETGs
and using only those objects detected in both
dust and gas. The probability that the two samples are drawn from
the same population ($p_{\rm KS}$) is $0.05$ and $0.89$ for the
global and central $M_{\rm dust}/M_{\rm gas}$, respectively, again
suggesting that the difference between Virgo Cluster ETGs and
group and field ETGs is more significant for the global
$M_{\rm dust}/M_{\rm gas}$ than the central $M_{\rm dust}/M_{\rm gas}$.
This thus indicates again that the central cold dense gas is more resilient
against ram pressure stripping than the diffuse \ion{H}{i} envelopes,
as suggested by \citet{you11} to explain the CO observations of the
ATLAS$^{\rm 3D}$ ETGs. Fig.~\ref{fig:density} also shows that the ETGs
with the lowest $M_{\rm dust}/M_{\rm gas}$ are in groups or the field,
where removal of the cold ISM of galaxies rarely occurs and the
$M_{\rm dust}/M_{\rm gas}$ may be diluted by external gas accretion
(see Sect.~\ref{sec:dust-gas}).

Since the cold dense ISM (H$_2$) is more important than the diffuse
atomic gas (\ion{H}{i}) for star formation, the above result suggests
that fast-rotating Virgo Cluster ETGs should form stars analogously to
group and field ETGs. As shown in the bottom panels of
Fig.~\ref{fig:density}, the $L_{\rm X}/M_{\rm dust}$ of fast-rotating ETGs
(a proxy for the star formation efficiency) indeed does not correlate with
$\rho_{10}$, nor does $L_{\rm X}/M_\star$ (a proxy for the specific star
formation rate), suggesting that cluster environments do not efficiently
suppress star formation in ETGs.

The bottom-right panel of Fig.~\ref{fig:density} also shows that
slow-rotating ETGs generally have higher $L_{\rm X}/M_\star$ than
fast-rotating ETGs. However, as slow-rotating ETGs tend to have
little current star formation \citep[e.g.][]{sha10,kok17}, their
X-ray luminosities are probably poor tracers of their SFRs and
instead likely reflect their substantial diffuse X-ray plasma.

To investigate potential environmental effects of groups on the cold
ISM of ETGs, we divide our
sample ETGs at $\log(\rho_{10}/{\rm Mpc}^3)<-0.4$ into group
or field ETGs using the galaxy group catalogue of \citet{gar93}, that is
complete to an apparent $B$-band magnitude of $14$ and a recession
velocity of $5500$~km~s$^{-1}$. We find that $164$
ETGs (out of $202$) meet these criteria, and $107$ of those belong to
a galaxy group. Figure~\ref{fig:group} is as Fig.~\ref{fig:density}, but for
group and field ETGs only. The top panels
show that group ETGs tend to have lower $M_{\rm dust}/M_\star$ and
$M_{\rm gas}/M_\star$ than field ETGs, although the statistical
significance of this is weak in both cases according to a K-S test
using FIR- and gas-detected ETGs ($p_{\rm KS}>0.1$). For
$M_{\rm dust}/M_\star$, we also perform Gehan's generalised Wilcoxon
test implemented in the ASURV package \citep{fei85,iso86}, including
non-FIR-detected ETGs. We find a marginal difference between group
and field ETGs ($p_{\rm GW}=0.03$), again suggesting that group
ETGs provide a slightly harsher environment for dust, possibly due
to denser X-ray plasma. However, it is difficult to support that
suggestion from the bottom panels of Fig.~\ref{fig:group}, as the
number of X-ray detected field ETGs is very small. This trend
should therefore be verified with more sensitive and numerous
observations of both dust and X-rays.

As Gehan's generalised Wilcoxon test cannot handle upper and lower
limits simultaneously, we cannot perform an analogous test for
$M_{\rm gas}/M_\star$, the global $M_{\rm dust}/M_{\rm gas}$ or
the central $M_{\rm dust}/M_{\rm gas}$.

In each galaxy group, \citet{gar93} also identified the
optically-brightest galaxy, likely to be located near the centre of
its group. We indicate them with additional larger open circles in
Fig.~\ref{fig:group}. The top panels show that the
optically-brightest group ETGs tend to have lower
$M_{\rm dust}/M_\star$ and $M_{\rm gas}/M_\star$ than other group
ETGs, suggesting that X-ray emitting IGM are denser in group cores
than at larger radii \citep[as expected; e.g.][]{ebe94}, and thus
heating and/or destruction of cold ISM may proceed more efficiently
in galaxies located there. However, although the number of galaxies
available is rather small, the bottom-right panel of
Fig.~\ref{fig:group} shows that the optically-brightest ETGs have
$L_{\rm X}/M_\star$ similar to those of other group ETGs. Again, the
above trend should therefore be verified with more sensitive and
numerous dust and X-ray observations.

%-----------------------------------------------------------------------

\section{Conclusion}
\label{sec:con}

We have systematically investigated the properties of the dust in the
cold and hot gas phases of the $260$ ETGs from the ATLAS$^{\rm 3D}$
survey, primarily relying on AKARI measurements. We found that dust is
prevalent in ETGs, with a detection rate of ${\approx}45{\%}$ in the
FIR, roughly the same for fast- and slow-rotating ETGs. Using SED fits
from our previous study, we derived the dust temperatures and masses
of each ETG. Our sample ETGs tend to have higher cold dust
temperatures $T_{\rm cold}$ than those of LTGs, in agreement with the
suggestion that ETGs possess intense stellar radiation fields due to
their dense populations of old stars. We found no correlation between
the total dust masses $M_{\rm dust}$ and total stellar masses
$M_\star$, implying that stellar mass loss is not a significant source
of dust in ETGs. Fast-rotating ETGs have higher global dust-to-stellar
mass ratios $M_{\rm dust}/M_\star$ than slow-rotating ETGs, consistent
with the fact that molecular gas and current star formation are
preferentially associated with the former.

The global dust-to-gas mass ratios $M_{\rm dust}/M_{\rm gas}$ of our
sample ETGs tend to be smaller than those of LTGs, which we have
argued is due to extended but dust-poor \ion{H}{i} envelopes. When the
\ion{H}{i} masses $M_\ion{H}{i}$ are estimated for the central regions
of the ETGs only, however, our ETGs show central
$M_{\rm dust}/M_{\rm gas}$ similar to those of LTGs, suggesting that
their central cold dense ISM have metallicities similar to those of
LTGs. This result can naturally explain why ETGs have star formation
efficiencies similar to those of LTGs. We also found some evidence
that ETGs with disrupted CO morphologies have lower central
$M_{\rm dust}/M_{\rm gas}$, suggesting that external gas accretion may
dilute the $M_{\rm dust}/M_{\rm gas}$ of some ETGs.

Adopting X-ray luminosities $L_{\rm X}$ that comprise the diffuse
X-ray gas only, we find that slow-rotating ETGs
have lower $M_{\rm dust}/M_\star$ than fast-rotating ETGs,
especially at higher $L_{\rm X}/M_\star$.
In contrast, fast-rotating ETGs show a correlation
between $M_{\rm dust}/M_\star$ and $L_{\rm X}/M_\star$, that may be
caused by recent star formation activity through supernova explosions
and/or YSOs. These results support currently favoured scenarios of ETG
evolution: fast-rotating ETGs still harbour star formation, while
slow-rotating ETGs have X-ray plasma-dominated ISM due to gas heating
through mergers and/or secular processes (e.g.\ stellar motions and
gravitational potentials).

We have also studied the effect of environment on the cold ISM of
ETGs. Virgo Cluster ETGs have higher global $M_{\rm dust}/M_{\rm gas}$
than group and field ETGs, suggesting that ram pressure stripping more easily
affects the diffuse \ion{H}{i} envelopes than the dust.
$M_{\rm dust}/M_{\rm gas}$ has no dependence on local galaxy density
when the central $M_{\rm dust}/M_{\rm gas}$ are used instead, nor does
the $L_{\rm X}/M_{\rm dust}$ of fast-rotating ETGs. These results
indicate that the central cold dense gas, essential for star formation
in ETGs, is relatively resilient against ram pressure stripping. Star
formation in cluster ETGs may therefore not easily be suppressed.

%--------------------------------------------------------------------
%\clearpage
%--------------------------------------------------------------------

%--------------------------------------------------------------------
%\clearpage
%--------------------------------------------------------------------

\begin{acknowledgements}
We thank the referee for carefully reading our manuscript and giving us
helpful comments. This research is based on observations with AKARI,
a JAXA project with the participation of ESA.
\end{acknowledgements}

% WARNING
%-------------------------------------------------------------------
% Please note that we have included the references to the file aa.dem in
% order to compile it, but we ask you to:
%
% - use BibTeX with the regular commands:

   \bibliographystyle{aa} % style aa.bst
   \bibliography{ref} % your references Yourfile.bib
%
% - join the .bib files when you upload your source files
%-------------------------------------------------------------------

\begin{appendix}

%\clearpage

\section{Combined measurements for the ATLAS$^{\rm 3D}$ galaxies}

\longtab[1]{
\begin{landscape}
\begin{longtable}{lrrrrrrrr}
  \caption{Combined measurements for the ATLAS$^{\rm 3D}$ galaxies.} \\
  \hline\hline
  Galaxy & $T_{\rm warm}$ & $T_{\rm cold}$ & log~$M_{\rm warm}$ & log~$M_{\rm cold}$ & log~$M_{\rm H_2}$\tablefootmark{a} & log~$M_{\rm global\ \ion{H}{i}}$\tablefootmark{b} & log~$M_{\rm central\ \ion{H}{i}}$\tablefootmark{c} & $L_{\rm X}$\tablefootmark{d}\\
  & (K) & (K) & ($M_\sun$) & ($M_\sun$) & ($M_\sun$) & ($M_\sun$) & ($M_\sun$) & ($10^{40}$~erg~s$^{-1}$)\\
  \hline
  \endfirsthead
  \caption{continued.}\\
  \hline\hline
  Galaxy & $T_{\rm warm}$ & $T_{\rm cold}$ & log~$M_{\rm warm}$ & log~$M_{\rm cold}$ & log~$M_{\rm H_2}$\tablefootmark{a} & log~$M_{\rm global\ \ion{H}{i}}$\tablefootmark{b} & log~$M_{\rm central\ \ion{H}{i}}$\tablefootmark{c} & $L_{\rm X}$\tablefootmark{d}\\
  & (K) & (K) & ($M_\sun$) & ($M_\sun$) & ($M_\sun$) & ($M_\sun$) & ($M_\sun$) & ($10^{40}$~erg~s$^{-1}$)\\
  \hline
  \endhead
  \hline
  \endfoot
  \hline
  \endlastfoot
IC0560 & ...\tablefootmark{e} & ...\tablefootmark{e} & 1.84 $\pm$ 0.17 & 5.45 $\pm$ 0.15 & $<$7.67 & ... & ... & ... \\
IC0598 & ...\tablefootmark{e} & ...\tablefootmark{e} & 1.70 $\pm$ 0.19 & 5.54 $\pm$ 0.12 & $<$8.02 & $<$7.45 & $<$7.06 & ... \\
IC0676 & 86.3 $\pm$ 2.0 & 28.8 $\pm$ 0.6 & 3.30 $\pm$ 0.05 & 6.09 $\pm$ 0.07 & 8.63 $\pm$ 0.02 & ... & ... & ... \\
IC0719 & 73.1 $\pm$ 2.2 & 24.5 $\pm$ 1.0 & 2.94 $\pm$ 0.12 & 6.47 $\pm$ 0.08 & 8.26 $\pm$ 0.04 & ... & ... & ... \\
IC0782 & ...\tablefootmark{e} & ...\tablefootmark{e} & 0.86 $\pm$ 0.84 & 5.60 $\pm$ 0.14 & $<$7.92 & ... & ... & ... \\
IC1024 & 101.3 $\pm$ 2.4 & 23.0 $\pm$ 0.3 & 2.43 $\pm$ 0.05 & 6.97 $\pm$ 0.06 & 8.61 $\pm$ 0.02 & ... & ... & 0.095$^{\rm +0.024}_{\rm -0.032}$ \\
IC3631 & ...\tablefootmark{e} & ...\tablefootmark{e} & 1.58 $\pm$ 0.26 & $<$5.64\tablefootmark{f} & $<$7.94 & $<$7.71 & $<$7.34 & ... \\
NGC0448 & ...\tablefootmark{e} & ...\tablefootmark{e} & 1.15 $\pm$ 0.50 & 3.99 $\pm$ 3.34 & $<$7.74 & ... & ... & ... \\
NGC0474 & ...\tablefootmark{e} & ...\tablefootmark{e} & 1.54 $\pm$ 0.27 & 4.66 $\pm$ 1.14 & $<$7.68 & ... & ... & ... \\
NGC0502 & ...\tablefootmark{e} & ...\tablefootmark{e} & 0.84 $\pm$ 0.47 & 3.99 $\pm$ 4.73 & $<$7.88 & ... & ... & ... \\
NGC0509 & ...\tablefootmark{e} & ...\tablefootmark{e} & 1.50 $\pm$ 0.20 & $<$5.32\tablefootmark{f} & 7.48 $\pm$ 0.12 & ... & ... & ... \\
NGC0516 & ...\tablefootmark{e} & ...\tablefootmark{e} & 0.55 $\pm$ 1.48 & 5.50 $\pm$ 0.17 & $<$7.82 & ... & ... & ... \\
NGC0524 & 74.1 $\pm$ 3.1 & 21.8 $\pm$ 0.5 & 2.69 $\pm$ 0.16 & 6.60 $\pm$ 0.08 & 7.97 $\pm$ 0.05 & ... & ... & ... \\
NGC0525 & ...\tablefootmark{e} & ...\tablefootmark{e} & 1.12 $\pm$ 0.27 & $<$5.51\tablefootmark{f} & $<$7.75 & ... & ... & ... \\
NGC0661 & ...\tablefootmark{e} & ...\tablefootmark{e} & 1.37 $\pm$ 0.33 & 5.45 $\pm$ 0.17 & $<$7.75 & $<$7.37 & $<$6.99 & ... \\
NGC0680 & ...\tablefootmark{e} & ...\tablefootmark{e} & 1.85 $\pm$ 0.17 & 5.57 $\pm$ 0.18 & $<$7.87 & 9.47 & 7.65 $\pm$ 0.05 & ... \\
NGC1023 & ...\tablefootmark{e} & ...\tablefootmark{e} & 1.85 $\pm$ 0.16 & 3.31 $\pm$ 1.92 & $<$6.79 & 9.29 & 6.84 $\pm$ 0.02 & 0.659$^{\rm +0.725}_{\rm -0.348}$ \\
NGC1222 & 83.6 $\pm$ 1.3 & 32.7 $\pm$ 0.7 & 4.08 $\pm$ 0.05 & 6.67 $\pm$ 0.07 & 9.07 $\pm$ 0.01 & ... & ... & ... \\
NGC1248 & ...\tablefootmark{e} & ...\tablefootmark{e} & 1.43 $\pm$ 0.26 & 5.07 $\pm$ 0.31 & $<$7.68 & ... & ... & ... \\
NGC1266 & 76.3 $\pm$ 0.7 & 31.0 $\pm$ 0.6 & 3.92 $\pm$ 0.05 & 6.75 $\pm$ 0.06 & 9.28 $\pm$ 0.01 & ... & ... & 0.288$^{\rm +0.071}_{\rm -0.049}$ \\
NGC1289 & 77.3 $\pm$ 4.1 & 24.9 $\pm$ 0.8 & 1.78 $\pm$ 0.28 & 6.24 $\pm$ 0.09 & $<$7.89 & ... & ... & ... \\
NGC1665 & ...\tablefootmark{e} & ...\tablefootmark{e} & 0.93 $\pm$ 1.41 & $<$5.68\tablefootmark{f} & $<$7.95 & ... & ... & ... \\
NGC2549 & ...\tablefootmark{e} & ...\tablefootmark{e} & 1.35 $\pm$ 0.19 & $<$4.58\tablefootmark{f} & $<$7.06 & $<$6.51 & $<$6.12 & ... \\
NGC2577 & 54.8 $\pm$ 3.1 & 15.2 $\pm$ 1.0 & 3.54 $\pm$ 0.16 & 7.81 $\pm$ 0.24 & $<$7.71 & $<$7.35 & $<$6.96 & ... \\
NGC2592 & ...\tablefootmark{e} & ...\tablefootmark{e} & 1.26 $\pm$ 0.28 & $<$5.30\tablefootmark{f} & $<$7.54 & $<$7.18 & $<$6.8 & ... \\
NGC2679 & ...\tablefootmark{e} & ...\tablefootmark{e} & 1.71 $\pm$ 0.19 & 5.36 $\pm$ 0.15 & $<$7.87 & $<$7.35 & $<$6.97 & ... \\
NGC2685 & 52.9 $\pm$ 1.3 & 13.9 $\pm$ 0.5 & 3.88 $\pm$ 0.12 & 7.35 $\pm$ 0.20 & 7.29 $\pm$ 0.08 & 9.33 & 7.36 $\pm$ 0.02 & ... \\
NGC2699 & ...\tablefootmark{e} & ...\tablefootmark{e} & 1.03 $\pm$ 0.34 & 4.75 $\pm$ 0.39 & $<$7.54 & ... & ... & ... \\
NGC2764 & 119.2 $\pm$ 2.8 & 25.4 $\pm$ 0.6 & 2.41 $\pm$ 0.06 & 7.11 $\pm$ 0.06 & 9.19 $\pm$ 0.02 & 9.28 & 8.91 $\pm$ 0.01 & ... \\
NGC2768 & ...\tablefootmark{e} & ...\tablefootmark{e} & 2.02 $\pm$ 0.21 & 5.43 $\pm$ 0.12 & 7.64 $\pm$ 0.07 & 7.81 & $<$6.61 & 1.249$^{\rm +0.246}_{\rm -0.183}$ \\
NGC2824 & 115.8 $\pm$ 3.3 & 28.6 $\pm$ 0.8 & 1.81 $\pm$ 0.09 & 6.21 $\pm$ 0.08 & 8.65 $\pm$ 0.03 & 7.59 & 7.45 $\pm$ 0.08 & ... \\
NGC2852 & ...\tablefootmark{e} & ...\tablefootmark{e} & 0.91 $\pm$ 0.40 & 4.41 $\pm$ 0.89 & $<$7.68 & $<$7.27 & $<$6.9 & ... \\
NGC2859 & ...\tablefootmark{e} & ...\tablefootmark{e} & 1.67 $\pm$ 0.30 & 5.72 $\pm$ 0.09 & $<$7.61 & 8.46 & $<$6.85 & ... \\
NGC2880 & 79.2 $\pm$ 8.9 & 18.8 $\pm$ 0.8 & 1.10 $\pm$ 0.48 & 6.06 $\pm$ 0.13 & $<$7.44 & $<$7.03 & $<$6.65 & ... \\
NGC2950 & ...\tablefootmark{e} & ...\tablefootmark{e} & 1.44 $\pm$ 0.24 & 4.32 $\pm$ 0.21 & $<$7.12 & $<$6.69 & $<$6.31 & ... \\
NGC2962 & 60.9 $\pm$ 1.8 & 16.3 $\pm$ 0.7 & 3.30 $\pm$ 0.18 & 7.38 $\pm$ 0.10 & $<$7.85 & ... & ... & ... \\
NGC3032 & 126.5 $\pm$ 3.0 & 26.8 $\pm$ 0.7 & 1.44 $\pm$ 0.06 & 6.09 $\pm$ 0.07 & 8.41 $\pm$ 0.01 & 8.04 & 7.8 $\pm$ 0.01 & ... \\
NGC3073 & ...\tablefootmark{e} & ...\tablefootmark{e} & 1.70 $\pm$ 0.17 & 4.74 $\pm$ 0.53 & 7.52 $\pm$ 0.07 & 8.56 & 8.01 $\pm$ 0.02 & ... \\
NGC3098 & ...\tablefootmark{e} & ...\tablefootmark{e} & 1.28 $\pm$ 0.33 & $<$5.14\tablefootmark{f} & $<$7.47 & $<$7.12 & $<$6.73 & ... \\
NGC3156 & 69.6 $\pm$ 1.9 & 21.1 $\pm$ 1.0 & 2.14 $\pm$ 0.18 & 6.03 $\pm$ 0.11 & 7.67 $\pm$ 0.09 & ... & ... & ... \\
NGC3182 & 116.9 $\pm$ 5.9 & 24.8 $\pm$ 0.8 & 1.33 $\pm$ 0.13 & 5.86 $\pm$ 0.10 & 8.33 $\pm$ 0.05 & 6.92 & 6.93 $\pm$ 0.16 & ... \\
NGC3193 & ...\tablefootmark{e} & ...\tablefootmark{e} & 1.69 $\pm$ 0.35 & 4.60 $\pm$ 0.94 & $<$7.91 & 8.19 & $<$7.07 & ... \\
NGC3230 & ...\tablefootmark{e} & ...\tablefootmark{e} & 1.72 $\pm$ 0.37 & 5.49 $\pm$ 0.18 & $<$8.00 & $<$7.71 & $<$7.33 & ... \\
NGC3245 & 111.9 $\pm$ 3.1 & 28.4 $\pm$ 0.5 & 1.78 $\pm$ 0.07 & 5.89 $\pm$ 0.07 & 7.27 $\pm$ 0.12 & $<$7 & $<$6.61 & ... \\
NGC3248 & ...\tablefootmark{e} & ...\tablefootmark{e} & $<$1.27\tablefootmark{f} & $<$5.11\tablefootmark{f} & $<$7.55 & $<$7.22 & $<$6.84 & ... \\
NGC3301 & 54.1 $\pm$ 1.4 & 16.2 $\pm$ 0.7 & 3.85 $\pm$ 0.14 & 7.10 $\pm$ 0.15 & $<$7.46 & $<$7.13 & $<$6.75 & ... \\
NGC3377 & ...\tablefootmark{e} & ...\tablefootmark{e} & 1.37 $\pm$ 0.20 & 3.62 $\pm$ 1.05 & $<$6.96 & $<$6.52 & $<$6.14 & 0.0032$^{\rm +0.0064}_{\rm -0.0032}$ \\
NGC3379 & ...\tablefootmark{e} & ...\tablefootmark{e} & 1.54 $\pm$ 0.23 & $<$4.23\tablefootmark{f} & $<$6.72 & $<$6.49 & $<$6.11 & 0.010$^{\rm +0.003}_{\rm -0.0025}$ \\
NGC3384 & ...\tablefootmark{e} & ...\tablefootmark{e} & 1.46 $\pm$ 0.23 & $<$4.43\tablefootmark{f} & $<$7.11 & 7.25 & $<$6.19 & 0.007$^{\rm +0.004}_{\rm -0.004}$ \\
NGC3400 & ...\tablefootmark{e} & ...\tablefootmark{e} & 1.07 $\pm$ 0.30 & 4.40 $\pm$ 0.63 & $<$7.63 & $<$7.19 & $<$6.81 & ... \\
NGC3412 & ...\tablefootmark{e} & ...\tablefootmark{e} & 0.83 $\pm$ 0.49 & $<$4.51\tablefootmark{f} & $<$6.96 & $<$6.55 & $<$6.17 & ... \\
NGC3414 & 54.9 $\pm$ 1.5 & 14.2 $\pm$ 0.6 & 3.82 $\pm$ 0.13 & 7.30 $\pm$ 0.22 & $<$7.19 & 8.28 & $<$7.7 & ... \\
NGC3457 & ...\tablefootmark{e} & ...\tablefootmark{e} & 0.95 $\pm$ 0.27 & 5.01 $\pm$ 0.15 & $<$7.35 & 8.07 & 6.95 $\pm$ 0.07 & ... \\
NGC3489 & 77.3 $\pm$ 1.9 & 27.3 $\pm$ 0.6 & 2.20 $\pm$ 0.12 & 5.49 $\pm$ 0.07 & 7.20 $\pm$ 0.06 & 6.87 & 6.53 $\pm$ 0.03 & ... \\
NGC3499 & ...\tablefootmark{e} & ...\tablefootmark{e} & 1.35 $\pm$ 0.19 & 4.89 $\pm$ 0.17 & $<$7.62 & 6.81 & 6.77 $\pm$ 0.14 & ... \\
NGC3522 & ...\tablefootmark{e} & ...\tablefootmark{e} & $<$0.29\tablefootmark{f} & 5.01 $\pm$ 0.21 & $<$7.28 & 8.47 & $<$7.48 & ... \\
NGC3530 & ...\tablefootmark{e} & ...\tablefootmark{e} & 1.02 $\pm$ 0.35 & 5.07 $\pm$ 0.18 & $<$7.78 & $<$7.37 & $<$6.98 & ... \\
NGC3595 & ...\tablefootmark{e} & ...\tablefootmark{e} & 0.90 $\pm$ 0.73 & 5.73 $\pm$ 0.12 & $<$7.84 & $<$7.43 & $<$7.04 & ... \\
NGC3599 & ...\tablefootmark{e} & ...\tablefootmark{e} & 1.80 $\pm$ 0.16 & 5.27 $\pm$ 0.12 & 7.36 $\pm$ 0.08 & $<$7.03 & $<$6.64 & 0.015$^{\rm +0.016}_{\rm -0.007}$ \\
NGC3607 & 102.0 $\pm$ 5.4 & 26.8 $\pm$ 0.5 & 1.76 $\pm$ 0.17 & 6.12 $\pm$ 0.07 & 8.42 $\pm$ 0.05 & $<$6.92 & $<$6.53 & 0.746$^{\rm +0.106}_{\rm -0.084}$ \\
NGC3608 & ...\tablefootmark{e} & ...\tablefootmark{e} & 1.55 $\pm$ 0.29 & 2.42 $\pm$ 106.66 & $<$7.58 & 7.16 & $<$6.53 & 0.358$^{\rm +0.075}_{\rm -0.060}$ \\
NGC3610 & ...\tablefootmark{e} & ...\tablefootmark{e} & 1.77 $\pm$ 0.17 & 3.48 $\pm$ 5.17 & $<$7.40 & $<$7.02 & $<$6.63 & ... \\
NGC3613 & ...\tablefootmark{e} & ...\tablefootmark{e} & 1.37 $\pm$ 0.60 & $<$5.30\tablefootmark{f} & $<$7.66 & $<$7.28 & $<$6.9 & ... \\
NGC3619 & 117.5 $\pm$ 7.4 & 40.1 $\pm$ 1.2 & 1.18 $\pm$ 0.16 & 4.88 $\pm$ 0.08 & 8.28 $\pm$ 0.05 & 9 & 8.25 $\pm$ 0.01 & ... \\
NGC3626 & 84.0 $\pm$ 1.5 & 25.2 $\pm$ 0.5 & 2.62 $\pm$ 0.08 & 6.26 $\pm$ 0.07 & 8.21 $\pm$ 0.04 & 8.94 & 7.8 $\pm$ 0.02 & ... \\
NGC3630 & ...\tablefootmark{e} & ...\tablefootmark{e} & 1.27 $\pm$ 0.30 & 4.56 $\pm$ 0.85 & $<$7.60 & ... & ... & ... \\
NGC3640 & ...\tablefootmark{e} & ...\tablefootmark{e} & 2.14 $\pm$ 0.20 & $<$5.14\tablefootmark{f} & $<$7.59 & ... & ... & ... \\
NGC3648 & ...\tablefootmark{e} & ...\tablefootmark{e} & 1.44 $\pm$ 0.25 & $<$5.25\tablefootmark{f} & $<$7.77 & $<$7.38 & $<$6.99 & ... \\
NGC3658 & ...\tablefootmark{e} & ...\tablefootmark{e} & 1.87 $\pm$ 0.19 & 5.29 $\pm$ 0.17 & $<$7.82 & $<$7.42 & $<$7.04 & ... \\
NGC3665 & 97.1 $\pm$ 3.2 & 23.9 $\pm$ 0.5 & 2.42 $\pm$ 0.11 & 6.93 $\pm$ 0.07 & 8.91 $\pm$ 0.02 & $<$7.43 & $<$7.05 & 1.919$^{\rm +0.385}_{\rm -0.287}$ \\
NGC3674 & ...\tablefootmark{e} & ...\tablefootmark{e} & $<$1.59\tablefootmark{f} & $<$5.22\tablefootmark{f} & $<$7.78 & $<$7.41 & $<$7.02 & ... \\
NGC3694 & 101.5 $\pm$ 5.4 & 39.2 $\pm$ 1.2 & 1.67 $\pm$ 0.16 & 5.25 $\pm$ 0.07 & $<$7.91 & $<$7.49 & $<$7.11 & ... \\
NGC3757 & ...\tablefootmark{e} & ...\tablefootmark{e} & 0.35 $\pm$ 0.99 & $<$4.85\tablefootmark{f} & $<$7.48 & $<$7.1 & $<$6.72 & ... \\
NGC3796 & 65.7 $\pm$ 4.1 & 23.3 $\pm$ 1.2 & 2.16 $\pm$ 0.18 & 5.43 $\pm$ 0.12 & $<$7.51 & $<$7.1 & $<$6.72 & ... \\
NGC3838 & ...\tablefootmark{e} & ...\tablefootmark{e} & 0.89 $\pm$ 0.38 & 5.07 $\pm$ 0.13 & $<$7.53 & 8.38 & $<$7.23 & ... \\
NGC3941 & ...\tablefootmark{e} & ...\tablefootmark{e} & 1.57 $\pm$ 0.16 & 4.63 $\pm$ 0.12 & $<$6.89 & 8.73 & $<$6.17 & ... \\
NGC3945 & 55.9 $\pm$ 1.4 & 28.4 $\pm$ 1.1 & 3.73 $\pm$ 0.15 & 5.31 $\pm$ 0.13 & $<$7.50 & 8.85 & $<$6.73 & ... \\
NGC3998 & 123.3 $\pm$ 3.0 & 27.5 $\pm$ 0.7 & 1.19 $\pm$ 0.07 & 4.94 $\pm$ 0.08 & $<$7.06 & 8.45 & 7.42 $\pm$ 0.02 & ... \\
NGC4026 & ...\tablefootmark{e} & ...\tablefootmark{e} & 1.15 $\pm$ 0.40 & 3.53 $\pm$ 1.16 & $<$6.99 & 8.5 & $<$7.14 & ... \\
NGC4036 & 105.4 $\pm$ 6.6 & 44.6 $\pm$ 1.7 & 1.47 $\pm$ 0.17 & 4.59 $\pm$ 0.07 & 8.13 $\pm$ 0.04 & 8.41 & $<$6.8 & ... \\
NGC4078 & ...\tablefootmark{e} & ...\tablefootmark{e} & 1.49 $\pm$ 0.27 & 5.48 $\pm$ 0.16 & $<$7.98 & $<$7.64 & $<$7.26 & ... \\
NGC4111 & 88.2 $\pm$ 2.1 & 29.3 $\pm$ 0.8 & 1.94 $\pm$ 0.09 & 5.23 $\pm$ 0.08 & 7.22 $\pm$ 0.09 & 8.81 & 6.94 $\pm$ 0.04 & ... \\
NGC4119 & ...\tablefootmark{e} & ...\tablefootmark{e} & 1.89 $\pm$ 0.16 & 5.27 $\pm$ 0.10 & 7.88 $\pm$ 0.03 & $<$7.1 & $<$7.1 & ... \\
NGC4143 & ...\tablefootmark{e} & ...\tablefootmark{e} & 1.69 $\pm$ 0.17 & 4.62 $\pm$ 0.16 & $<$7.20 & $<$6.8 & $<$6.42 & ... \\
NGC4150 & 120.2 $\pm$ 4.8 & 39.7 $\pm$ 1.2 & 0.78 $\pm$ 0.10 & 4.50 $\pm$ 0.07 & 7.82 $\pm$ 0.03 & 6.26 & 6.04 $\pm$ 0.06 & ... \\
NGC4168 & ...\tablefootmark{e} & ...\tablefootmark{e} & 1.80 $\pm$ 0.25 & 5.33 $\pm$ 0.20 & $<$7.74 & $<$7.46 & $<$7.08 & ... \\
NGC4179 & ...\tablefootmark{e} & ...\tablefootmark{e} & $<$1.49\tablefootmark{f} & $<$4.94\tablefootmark{f} & $<$7.28 & ... & ... & ... \\
NGC4191 & ...\tablefootmark{e} & ...\tablefootmark{e} & 1.60 $\pm$ 0.26 & 5.13 $\pm$ 0.60 & $<$7.94 & ... & ... & ... \\
NGC4203 & 68.4 $\pm$ 2.6 & 26.9 $\pm$ 0.6 & 2.70 $\pm$ 0.15 & 5.43 $\pm$ 0.08 & 7.39 $\pm$ 0.05 & 9.15 & 7.03 $\pm$ 0.03 & 0.039$^{\rm +0.014}_{\rm -0.014}$ \\
NGC4215 & ...\tablefootmark{e} & ...\tablefootmark{e} & 1.49 $\pm$ 0.33 & $<$5.43\tablefootmark{f} & $<$7.83 & ... & ... & ... \\
NGC4233 & 55.9 $\pm$ 2.7 & 18.6 $\pm$ 0.8 & 3.57 $\pm$ 0.17 & 6.79 $\pm$ 0.19 & $<$7.89 & ... & ... & ... \\
NGC4249 & ...\tablefootmark{e} & ...\tablefootmark{e} & 0.11 $\pm$ 3.87 & $<$5.42\tablefootmark{f} & $<$7.97 & ... & ... & ... \\
NGC4251 & ...\tablefootmark{e} & ...\tablefootmark{e} & 1.73 $\pm$ 0.20 & 4.13 $\pm$ 0.56 & $<$7.11 & $<$6.97 & $<$6.58 & ... \\
NGC4255 & ...\tablefootmark{e} & ...\tablefootmark{e} & 0.67 $\pm$ 1.24 & 4.65 $\pm$ 0.70 & $<$7.78 & ... & ... & ... \\
NGC4259 & ...\tablefootmark{e} & ...\tablefootmark{e} & 0.67 $\pm$ 0.26 & 5.44 $\pm$ 0.28 & $<$7.97 & ... & ... & ... \\
NGC4261 & ...\tablefootmark{e} & ...\tablefootmark{e} & 2.51 $\pm$ 0.16 & 5.40 $\pm$ 0.20 & $<$7.68 & ... & ... & 4.261$^{\rm +0.038}_{\rm -0.038}$ \\
NGC4262 & ...\tablefootmark{e} & ...\tablefootmark{e} & 0.97 $\pm$ 0.41 & 4.21 $\pm$ 0.56 & $<$7.07 & 8.69 & $<$7.02 & ... \\
NGC4264 & ...\tablefootmark{e} & ...\tablefootmark{e} & 1.38 $\pm$ 0.29 & 5.54 $\pm$ 0.19 & $<$7.94 & ... & ... & ... \\
NGC4267 & ...\tablefootmark{e} & ...\tablefootmark{e} & 1.34 $\pm$ 0.26 & 3.85 $\pm$ 1.60 & $<$7.16 & $<$7.17 & $<$7.17 & ... \\
NGC4268 & 59.2 $\pm$ 4.0 & 18.4 $\pm$ 0.6 & 2.76 $\pm$ 0.29 & 6.72 $\pm$ 0.12 & $<$7.83 & ... & ... & ... \\
NGC4270 & ...\tablefootmark{e} & ...\tablefootmark{e} & 1.27 $\pm$ 0.61 & $<$5.45\tablefootmark{f} & $<$7.79 & ... & ... & ... \\
NGC4278 & 60.7 $\pm$ 1.4 & 29.7 $\pm$ 1.0 & 3.19 $\pm$ 0.15 & 5.11 $\pm$ 0.10 & $<$7.45 & 8.8 & 6.06 $\pm$ 0.09 & 0.088$^{\rm +0.007}_{\rm -0.012}$ \\
NGC4281 & ...\tablefootmark{e} & ...\tablefootmark{e} & 2.26 $\pm$ 0.15 & 5.60 $\pm$ 0.11 & $<$7.88 & ... & ... & ... \\
NGC4283 & ...\tablefootmark{e} & ...\tablefootmark{e} & 1.08 $\pm$ 0.20 & 4.47 $\pm$ 0.22 & 7.10 $\pm$ 0.09 & $<$6.36 & $<$5.97 & ... \\
NGC4324 & 86.3 $\pm$ 2.9 & 14.6 $\pm$ 0.4 & 1.73 $\pm$ 0.14 & 7.16 $\pm$ 0.10 & 7.69 $\pm$ 0.05 & ... & ... & ... \\
NGC4339 & ...\tablefootmark{e} & ...\tablefootmark{e} & 1.12 $\pm$ 0.35 & 4.61 $\pm$ 0.40 & $<$7.15 & ... & ... & ... \\
NGC4340 & ...\tablefootmark{e} & ...\tablefootmark{e} & 1.22 $\pm$ 0.34 & 4.75 $\pm$ 0.21 & $<$7.33 & $<$7.03 & $<$6.65 & ... \\
NGC4342 & ...\tablefootmark{e} & ...\tablefootmark{e} & 0.58 $\pm$ 0.35 & $<$4.77\tablefootmark{f} & $<$7.24 & ... & ... & 0.055$^{\rm +0.007}_{\rm -0.006}$ \\
NGC4346 & ...\tablefootmark{e} & ...\tablefootmark{e} & 1.35 $\pm$ 0.17 & $<$4.67\tablefootmark{f} & $<$7.12 & $<$6.66 & $<$6.27 & ... \\
NGC4350 & 58.5 $\pm$ 1.9 & 27.9 $\pm$ 0.9 & 3.08 $\pm$ 0.16 & 5.10 $\pm$ 0.10 & $<$7.18 & $<$6.88 & $<$6.5 & ... \\
NGC4365 & ...\tablefootmark{e} & ...\tablefootmark{e} & 2.19 $\pm$ 0.20 & $<$5.24\tablefootmark{f} & $<$7.62 & ... & ... & 0.544$^{\rm +0.041}_{\rm -0.040}$ \\
NGC4371 & ...\tablefootmark{e} & ...\tablefootmark{e} & 1.57 $\pm$ 0.22 & $<$5.00\tablefootmark{f} & $<$7.29 & $<$7.1 & $<$7.1 & ... \\
NGC4374 & 98.1 $\pm$ 4.4 & 23.4 $\pm$ 1.0 & 1.81 $\pm$ 0.14 & 5.88 $\pm$ 0.10 & $<$7.23 & $<$7.26 & $<$6.88 & 5.423$^{\rm +0.544}_{\rm -0.531}$ \\
NGC4377 & 54.4 $\pm$ 3.1 & 16.7 $\pm$ 0.4 & 2.50 $\pm$ 0.32 & 6.67 $\pm$ 0.10 & $<$7.26 & $<$7.16 & $<$7.16 & ... \\
NGC4379 & ...\tablefootmark{e} & ...\tablefootmark{e} & 0.72 $\pm$ 0.44 & 4.89 $\pm$ 0.18 & $<$7.19 & $<$7.04 & $<$7.04 & ... \\
NGC4382 & ...\tablefootmark{e} & ...\tablefootmark{e} & 2.29 $\pm$ 0.18 & $<$5.10\tablefootmark{f} & $<$7.39 & $<$6.97 & $<$6.59 & 1.378$^{\rm +0.251}_{\rm -0.259}$ \\
NGC4406 & ...\tablefootmark{e} & ...\tablefootmark{e} & 1.93 $\pm$ 0.26 & $<$5.19\tablefootmark{f} & $<$7.40 & 8 & $<$6.4 & 9.988$^{\rm +1.281}_{\rm -1.281}$ \\
NGC4417 & ...\tablefootmark{e} & ...\tablefootmark{e} & 1.46 $\pm$ 0.19 & 5.02 $\pm$ 0.13 & $<$7.22 & ... & ... & ... \\
NGC4425 & ...\tablefootmark{e} & ...\tablefootmark{e} & 0.62 $\pm$ 0.62 & $<$5.10\tablefootmark{f} & $<$7.20 & $<$6.71 & $<$6.33 & ... \\
NGC4429 & 71.1 $\pm$ 1.6 & 24.4 $\pm$ 0.6 & 3.00 $\pm$ 0.12 & 6.23 $\pm$ 0.08 & 8.05 $\pm$ 0.03 & $<$7.12 & $<$7.12 & ... \\
NGC4434 & ...\tablefootmark{e} & ...\tablefootmark{e} & 0.87 $\pm$ 0.47 & 4.74 $\pm$ 0.38 & $<$7.60 & ... & ... & ... \\
NGC4435 & 76.7 $\pm$ 1.7 & 19.7 $\pm$ 0.4 & 2.57 $\pm$ 0.13 & 6.81 $\pm$ 0.08 & 7.87 $\pm$ 0.04 & $<$7.23 & $<$7.23 & ... \\
NGC4458 & ...\tablefootmark{e} & ...\tablefootmark{e} & 0.34 $\pm$ 0.99 & 3.22 $\pm$ 6.39 & $<$7.31 & $<$6.91 & $<$6.53 & 0.004$^{\rm +0.010}_{\rm -0.004}$ \\
NGC4459 & 90.8 $\pm$ 3.5 & 26.7 $\pm$ 1.3 & 2.04 $\pm$ 0.11 & 6.09 $\pm$ 0.13 & 8.24 $\pm$ 0.02 & $<$6.91 & $<$6.53 & 0.181$^{\rm +0.018}_{\rm -0.017}$ \\
NGC4461 & ...\tablefootmark{e} & ...\tablefootmark{e} & 1.14 $\pm$ 0.33 & $<$4.64\tablefootmark{f} & $<$7.20 & $<$7.33 & $<$7.33 & ... \\
NGC4472 & ...\tablefootmark{e} & ...\tablefootmark{e} & 2.40 $\pm$ 0.18 & 2.58 $\pm$ 45.68 & $<$7.25 & ... & ... & 16.096$^{\rm +0.836}_{\rm -0.836}$ \\
NGC4474 & ...\tablefootmark{e} & ...\tablefootmark{e} & 1.13 $\pm$ 0.23 & $<$4.73\tablefootmark{f} & $<$7.16 & $<$7.08 & $<$7.09 & ... \\
NGC4476 & 123.9 $\pm$ 4.1 & 35.0 $\pm$ 0.8 & 0.62 $\pm$ 0.13 & 4.92 $\pm$ 0.08 & 8.05 $\pm$ 0.04 & ... & ... & ... \\
NGC4477 & 81.1 $\pm$ 3.9 & 28.6 $\pm$ 1.6 & 1.93 $\pm$ 0.17 & 5.34 $\pm$ 0.14 & 7.54 $\pm$ 0.06 & $<$6.95 & $<$6.56 & 0.740$^{\rm +0.027}_{\rm -0.027}$ \\
NGC4478 & ...\tablefootmark{e} & ...\tablefootmark{e} & 1.33 $\pm$ 0.24 & $<$5.01\tablefootmark{f} & $<$7.28 & ... & ... & ... \\
NGC4483 & 65.4 $\pm$ 9.7 & 20.5 $\pm$ 0.8 & 1.64 $\pm$ 0.47 & 5.70 $\pm$ 0.12 & $<$7.20 & ... & ... & ... \\
NGC4486 & ...\tablefootmark{e} & ...\tablefootmark{e} & 2.59 $\pm$ 0.13 & 5.00 $\pm$ 0.17 & $<$7.17 & ... & ... & ... \\
NGC4489 & ...\tablefootmark{e} & ...\tablefootmark{e} & 0.56 $\pm$ 0.42 & 4.43 $\pm$ 0.40 & $<$7.15 & $<$6.74 & $<$6.35 & ... \\
NGC4494 & ...\tablefootmark{e} & ...\tablefootmark{e} & 1.78 $\pm$ 0.20 & 4.70 $\pm$ 0.18 & $<$7.25 & $<$6.84 & $<$6.46 & 0.014$^{\rm +0.012}_{\rm -0.014}$ \\
NGC4503 & ...\tablefootmark{e} & ...\tablefootmark{e} & 1.36 $\pm$ 0.26 & $<$5.30\tablefootmark{f} & $<$7.22 & $<$7.14 & $<$7.15 & ... \\
NGC4521 & ...\tablefootmark{e} & ...\tablefootmark{e} & 2.16 $\pm$ 0.16 & 5.70 $\pm$ 0.11 & $<$7.97 & 7.75 & $<$7.18 & ... \\
NGC4526 & 102.1 $\pm$ 2.3 & 25.3 $\pm$ 0.6 & 2.02 $\pm$ 0.09 & 6.63 $\pm$ 0.06 & 8.59 $\pm$ 0.01 & ... & ... & 0.506$^{\rm +0.109}_{\rm -0.109}$ \\
NGC4528 & ...\tablefootmark{e} & ...\tablefootmark{e} & 0.99 $\pm$ 0.31 & 4.96 $\pm$ 0.14 & $<$7.15 & $<$7.18 & $<$7.19 & ... \\
NGC4546 & ...\tablefootmark{e} & ...\tablefootmark{e} & 1.69 $\pm$ 0.19 & 4.94 $\pm$ 0.13 & $<$6.97 & ... & ... & ... \\
NGC4550 & ...\tablefootmark{e} & ...\tablefootmark{e} & 1.02 $\pm$ 0.32 & 4.84 $\pm$ 0.20 & $<$7.24 & $<$6.89 & $<$6.5 & ... \\
NGC4551 & ...\tablefootmark{e} & ...\tablefootmark{e} & 0.87 $\pm$ 0.36 & $<$5.01\tablefootmark{f} & $<$7.24 & $<$7.39 & $<$7.39 & ... \\
NGC4552 & 66.1 $\pm$ 4.7 & 24.2 $\pm$ 1.7 & 2.75 $\pm$ 0.21 & 5.17 $\pm$ 0.15 & $<$7.28 & $<$6.87 & $<$6.48 & 2.207$^{\rm +0.110}_{\rm -0.811}$ \\
NGC4564 & ...\tablefootmark{e} & ...\tablefootmark{e} & 1.26 $\pm$ 0.31 & $<$5.01\tablefootmark{f} & $<$7.25 & $<$6.91 & $<$6.53 & 0.005$^{\rm +0.008}_{\rm -0.005}$ \\
NGC4570 & ...\tablefootmark{e} & ...\tablefootmark{e} & 1.65 $\pm$ 0.23 & 4.86 $\pm$ 0.31 & $<$7.47 & ... & ... & ... \\
NGC4578 & ...\tablefootmark{e} & ...\tablefootmark{e} & 1.43 $\pm$ 0.25 & 4.25 $\pm$ 0.64 & $<$7.20 & ... & ... & ... \\
NGC4596 & 63.8 $\pm$ 2.6 & 38.1 $\pm$ 2.1 & 2.72 $\pm$ 0.24 & 4.74 $\pm$ 0.08 & 7.31 $\pm$ 0.09 & $<$7.13 & $<$7.13 & 0.098$^{\rm +0.026}_{\rm -0.020}$ \\
NGC4608 & ...\tablefootmark{e} & ...\tablefootmark{e} & 1.09 $\pm$ 0.42 & $<$4.82\tablefootmark{f} & $<$7.30 & $<$7.22 & $<$7.22 & ... \\
NGC4612 & ...\tablefootmark{e} & ...\tablefootmark{e} & 1.26 $\pm$ 0.24 & $<$4.74\tablefootmark{f} & $<$7.20 & ... & ... & ... \\
NGC4621 & 117.4 $\pm$ 10.2 & 22.3 $\pm$ 0.7 & 0.71 $\pm$ 0.21 & 5.69 $\pm$ 0.10 & $<$7.13 & $<$6.86 & $<$6.48 & 0.053$^{\rm +0.103}_{\rm -0.026}$ \\
NGC4623 & ...\tablefootmark{e} & ...\tablefootmark{e} & 1.38 $\pm$ 0.19 & 3.65 $\pm$ 2.29 & $<$7.21 & ... & ... & ... \\
NGC4624 & ...\tablefootmark{e} & ...\tablefootmark{e} & 1.65 $\pm$ 0.29 & $<$5.03\tablefootmark{f} & $<$7.30 & ... & ... & ... \\
NGC4636 & ...\tablefootmark{e} & ...\tablefootmark{e} & 2.00 $\pm$ 0.19 & 5.16 $\pm$ 0.14 & $<$6.87 & ... & ... & 20.028$^{\rm +0.506}_{\rm -0.506}$ \\
NGC4638 & ...\tablefootmark{e} & ...\tablefootmark{e} & 1.10 $\pm$ 0.55 & 4.02 $\pm$ 1.41 & $<$7.30 & $<$7.12 & $<$7.13 & ... \\
NGC4643 & 58.3 $\pm$ 1.8 & 24.0 $\pm$ 0.7 & 3.53 $\pm$ 0.16 & 5.93 $\pm$ 0.10 & 7.27 $\pm$ 0.12 & ... & ... & ... \\
NGC4660 & ...\tablefootmark{e} & ...\tablefootmark{e} & 1.42 $\pm$ 0.17 & $<$4.78\tablefootmark{f} & $<$7.19 & $<$6.88 & $<$6.5 & ... \\
NGC4684 & 79.0 $\pm$ 1.8 & 26.4 $\pm$ 1.0 & 2.62 $\pm$ 0.08 & 5.46 $\pm$ 0.09 & 7.21 $\pm$ 0.11 & ... & ... & ... \\
NGC4690 & ...\tablefootmark{e} & ...\tablefootmark{e} & 1.42 $\pm$ 0.31 & 5.51 $\pm$ 0.19 & $<$8.01 & ... & ... & ... \\
NGC4694 & 59.4 $\pm$ 2.2 & 27.6 $\pm$ 0.8 & 3.53 $\pm$ 0.16 & 5.71 $\pm$ 0.09 & 8.01 $\pm$ 0.03 & 8.21 & 7.13 $\pm$ 0.02 & ... \\
NGC4710 & 81.0 $\pm$ 1.5 & 26.7 $\pm$ 0.5 & 2.92 $\pm$ 0.07 & 6.48 $\pm$ 0.06 & 8.72 $\pm$ 0.01 & 6.84 & 6.71 $\pm$ 0.08 & 0.087$^{\rm +0.013}_{\rm -0.015}$ \\
NGC4733 & ...\tablefootmark{e} & ...\tablefootmark{e} & 0.87 $\pm$ 0.30 & $<$4.58\tablefootmark{f} & $<$7.28 & $<$7.12 & $<$7.12 & ... \\
NGC4753 & 115.7 $\pm$ 4.3 & 26.1 $\pm$ 1.1 & 1.86 $\pm$ 0.11 & 6.68 $\pm$ 0.16 & 8.55 $\pm$ 0.03 & ... & ... & ... \\
NGC4754 & ...\tablefootmark{e} & ...\tablefootmark{e} & 1.57 $\pm$ 0.26 & $<$4.78\tablefootmark{f} & $<$7.18 & $<$7.18 & $<$7.18 & ... \\
NGC4762 & ...\tablefootmark{e} & ...\tablefootmark{e} & 1.78 $\pm$ 0.37 & $<$5.08\tablefootmark{f} & $<$7.48 & $<$7.4 & $<$7.41 & ... \\
NGC4803 & ...\tablefootmark{e} & ...\tablefootmark{e} & $<$1.54\tablefootmark{f} & 5.52 $\pm$ 0.14 & $<$7.98 & ... & ... & ... \\
NGC5103 & ...\tablefootmark{e} & ...\tablefootmark{e} & 0.96 $\pm$ 0.40 & 3.81 $\pm$ 1.64 & $<$7.58 & 8.57 & 7.3 $\pm$ 0.04 & ... \\
NGC5173 & ...\tablefootmark{e} & ...\tablefootmark{e} & 2.11 $\pm$ 0.17 & 5.82 $\pm$ 0.09 & 8.28 $\pm$ 0.06 & 9.33 & 8.45 $\pm$ 0.01 & ... \\
NGC5198 & ...\tablefootmark{e} & ...\tablefootmark{e} & 1.74 $\pm$ 0.26 & $<$5.46\tablefootmark{f} & $<$7.89 & 8.49 & $<$6.98 & ... \\
NGC5273 & ...\tablefootmark{e} & ...\tablefootmark{e} & 2.07 $\pm$ 0.15 & 5.28 $\pm$ 0.10 & 7.31 $\pm$ 0.07 & $<$6.81 & $<$6.42 & ... \\
NGC5308 & ...\tablefootmark{e} & ...\tablefootmark{e} & 1.64 $\pm$ 0.35 & 5.05 $\pm$ 0.20 & $<$7.88 & $<$7.63 & $<$7.24 & ... \\
NGC5322 & 54.9 $\pm$ 1.7 & 33.5 $\pm$ 2.1 & 4.17 $\pm$ 0.15 & 5.03 $\pm$ 0.20 & $<$7.76 & $<$7.34 & $<$6.96 & ... \\
NGC5353 & 62.2 $\pm$ 2.0 & 25.7 $\pm$ 0.9 & 3.57 $\pm$ 0.16 & 6.09 $\pm$ 0.10 & $<$8.12 & $<$7.45 & $<$7.07 & ... \\
NGC5355 & ...\tablefootmark{e} & ...\tablefootmark{e} & 1.53 $\pm$ 0.21 & 4.39 $\pm$ 1.53 & $<$7.94 & $<$7.5 & $<$7.11 & ... \\
NGC5358 & ...\tablefootmark{e} & ...\tablefootmark{e} & 0.73 $\pm$ 0.67 & 5.30 $\pm$ 0.22 & $<$7.92 & $<$7.52 & $<$7.13 & ... \\
NGC5379 & ...\tablefootmark{e} & ...\tablefootmark{e} & 2.27 $\pm$ 0.15 & 5.37 $\pm$ 0.16 & 8.33 $\pm$ 0.04 & $<$7.36 & $<$6.97 & ... \\
NGC5422 & ...\tablefootmark{e} & ...\tablefootmark{e} & 1.66 $\pm$ 0.23 & 4.87 $\pm$ 0.31 & $<$7.78 & 7.87 & 7.43 $\pm$ 0.05 & 0.017$^{\rm +0.014}_{\rm -0.017}$ \\
NGC5473 & ...\tablefootmark{e} & ...\tablefootmark{e} & 1.96 $\pm$ 0.18 & 5.17 $\pm$ 0.19 & $<$7.85 & $<$7.4 & $<$7.02 & ... \\
NGC5475 & ...\tablefootmark{e} & ...\tablefootmark{e} & 1.57 $\pm$ 0.19 & 5.75 $\pm$ 0.19 & $<$7.72 & $<$7.28 & $<$6.89 & ... \\
NGC5485 & 69.4 $\pm$ 4.3 & 24.7 $\pm$ 3.1 & 2.43 $\pm$ 0.21 & 5.64 $\pm$ 0.10 & $<$7.60 & $<$7.17 & $<$6.79 & ... \\
NGC5493 & ...\tablefootmark{e} & ...\tablefootmark{e} & 2.00 $\pm$ 0.21 & 5.01 $\pm$ 0.68 & $<$7.98 & ... & ... & ... \\
NGC5500 & ...\tablefootmark{e} & ...\tablefootmark{e} & 0.97 $\pm$ 0.26 & $<$5.04\tablefootmark{f} & $<$7.82 & $<$7.36 & $<$6.97 & ... \\
NGC5557 & ...\tablefootmark{e} & ...\tablefootmark{e} & 2.03 $\pm$ 0.25 & $<$5.23\tablefootmark{f} & $<$7.92 & 8.57 & $<$7.16 & ... \\
NGC5574 & ...\tablefootmark{e} & ...\tablefootmark{e} & 1.38 $\pm$ 0.19 & $<$5.08\tablefootmark{f} & $<$7.51 & ... & ... & ... \\
NGC5576 & ...\tablefootmark{e} & ...\tablefootmark{e} & 1.70 $\pm$ 0.24 & 4.93 $\pm$ 0.22 & $<$7.60 & ... & ... & 0.036$^{\rm +0.042}_{\rm -0.020}$ \\
NGC5582 & ...\tablefootmark{e} & ...\tablefootmark{e} & 1.15 $\pm$ 0.33 & 4.64 $\pm$ 0.41 & $<$7.67 & 9.65 & $<$6.88 & ... \\
NGC5631 & 49.6 $\pm$ 1.2 & 21.2 $\pm$ 1.5 & 4.21 $\pm$ 0.13 & 5.72 $\pm$ 0.38 & $<$7.68 & 8.89 & 7.54 $\pm$ 0.03 & ... \\
NGC5638 & ...\tablefootmark{e} & ...\tablefootmark{e} & 1.69 $\pm$ 0.21 & 5.03 $\pm$ 0.20 & $<$7.60 & ... & ... & ... \\
NGC5687 & ...\tablefootmark{e} & ...\tablefootmark{e} & 1.43 $\pm$ 0.24 & $<$5.15\tablefootmark{f} & $<$7.64 & $<$7.32 & $<$6.94 & ... \\
NGC5770 & ...\tablefootmark{e} & ...\tablefootmark{e} & 0.90 $\pm$ 0.31 & $<$4.69\tablefootmark{f} & $<$7.34 & ... & ... & ... \\
NGC5813 & ...\tablefootmark{e} & ...\tablefootmark{e} & 2.08 $\pm$ 0.24 & 5.51 $\pm$ 0.18 & $<$7.69 & ... & ... & 68.87$^{\rm +1.958}_{\rm -1.953}$ \\
NGC5831 & ...\tablefootmark{e} & ...\tablefootmark{e} & 1.64 $\pm$ 0.21 & 5.10 $\pm$ 0.19 & $<$7.85 & ... & ... & ... \\
NGC5838 & 85.7 $\pm$ 5.1 & 22.9 $\pm$ 0.7 & 2.10 $\pm$ 0.14 & 6.25 $\pm$ 0.08 & $<$7.56 & ... & ... & ... \\
NGC5839 & ...\tablefootmark{e} & ...\tablefootmark{e} & 1.06 $\pm$ 0.28 & 4.73 $\pm$ 0.24 & $<$7.38 & ... & ... & ... \\
NGC5845 & ...\tablefootmark{e} & ...\tablefootmark{e} & 1.38 $\pm$ 0.20 & $<$5.59\tablefootmark{f} & $<$7.50 & ... & ... & ... \\
NGC5846 & ...\tablefootmark{e} & ...\tablefootmark{e} & 2.06 $\pm$ 0.22 & 5.26 $\pm$ 0.18 & $<$7.78 & ... & ... & 28.26$^{\rm +1.170}_{\rm -1.163}$ \\
NGC5854 & ...\tablefootmark{e} & ...\tablefootmark{e} & 1.70 $\pm$ 0.18 & 5.13 $\pm$ 0.23 & $<$7.60 & ... & ... & ... \\
NGC5864 & ...\tablefootmark{e} & ...\tablefootmark{e} & 1.66 $\pm$ 0.22 & $<$5.36\tablefootmark{f} & $<$7.74 & ... & ... & ... \\
NGC5866 & 120.3 $\pm$ 2.8 & 25.7 $\pm$ 0.6 & 1.37 $\pm$ 0.07 & 6.44 $\pm$ 0.06 & 8.47 $\pm$ 0.01 & 6.96 & 6.67 $\pm$ 0.06 & 0.473$^{\rm +0.307}_{\rm -0.197}$ \\
NGC5869 & ...\tablefootmark{e} & ...\tablefootmark{e} & 1.41 $\pm$ 0.24 & $<$5.09\tablefootmark{f} & $<$7.63 & ... & ... & ... \\
NGC6010 & ...\tablefootmark{e} & ...\tablefootmark{e} & 1.66 $\pm$ 0.25 & $<$5.40\tablefootmark{f} & $<$7.78 & ... & ... & ... \\
NGC6014 & 74.8 $\pm$ 1.9 & 25.9 $\pm$ 0.8 & 3.44 $\pm$ 0.09 & 6.39 $\pm$ 0.08 & 8.77 $\pm$ 0.02 & ... & ... & ... \\
NGC6017 & ...\tablefootmark{e} & ...\tablefootmark{e} & 1.92 $\pm$ 0.14 & 5.34 $\pm$ 0.17 & $<$7.73 & ... & ... & ... \\
NGC6149 & ...\tablefootmark{e} & ...\tablefootmark{e} & 1.58 $\pm$ 0.24 & 5.67 $\pm$ 0.11 & $<$7.90 & $<$7.56 & $<$7.18 & ... \\
NGC6278 & ...\tablefootmark{e} & ...\tablefootmark{e} & 2.01 $\pm$ 0.19 & 5.83 $\pm$ 0.11 & $<$7.98 & $<$7.67 & $<$7.28 & ... \\
NGC6547 & ...\tablefootmark{e} & ...\tablefootmark{e} & 1.35 $\pm$ 0.46 & 5.56 $\pm$ 0.20 & $<$8.00 & $<$7.63 & $<$7.25 & ... \\
NGC6548 & ...\tablefootmark{e} & ...\tablefootmark{e} & 1.64 $\pm$ 0.20 & $<$5.13\tablefootmark{f} & $<$7.58 & $<$7.12 & $<$6.74 & ... \\
NGC6703 & ...\tablefootmark{e} & ...\tablefootmark{e} & 1.68 $\pm$ 0.25 & 5.14 $\pm$ 0.17 & $<$7.62 & $<$7.18 & $<$6.8 & ... \\
NGC6798 & ...\tablefootmark{e} & ...\tablefootmark{e} & 2.03 $\pm$ 0.17 & 4.15 $\pm$ 2.91 & 7.83 $\pm$ 0.10 & 9.38 & 8.1 $\pm$ 0.02 & ... \\
NGC7280 & ...\tablefootmark{e} & ...\tablefootmark{e} & 1.57 $\pm$ 0.21 & 5.00 $\pm$ 0.18 & $<$7.49 & 7.92 & 7.25 $\pm$ 0.05 & ... \\
NGC7332 & ...\tablefootmark{e} & ...\tablefootmark{e} & 1.89 $\pm$ 0.16 & 4.84 $\pm$ 0.28 & $<$7.41 & 6.62 & $<$6.7 & ... \\
NGC7457 & ...\tablefootmark{e} & ...\tablefootmark{e} & 1.01 $\pm$ 0.31 & 3.81 $\pm$ 0.59 & $<$6.96 & $<$6.61 & $<$6.22 & 0.003$^{\rm +0.008}_{\rm -0.003}$ \\
NGC7465 & 98.4 $\pm$ 1.7 & 24.6 $\pm$ 0.4 & 2.68 $\pm$ 0.05 & 6.90 $\pm$ 0.06 & 8.79 $\pm$ 0.02 & 9.98 & 8.64 $\pm$ 0.01 & ... \\
NGC7693 & ...\tablefootmark{e} & ...\tablefootmark{e} & $<$1.42\tablefootmark{f} & 5.39 $\pm$ 0.19 & $<$7.86 & ... & ... & ... \\
NGC7710 & ...\tablefootmark{e} & ...\tablefootmark{e} & 0.82 $\pm$ 0.67 & 5.22 $\pm$ 0.26 & $<$7.80 & ... & ... & ... \\
PGC016060 & 75.9 $\pm$ 1.9 & 27.2 $\pm$ 1.4 & 2.54 $\pm$ 0.15 & 6.08 $\pm$ 0.09 & 8.26 $\pm$ 0.06 & ... & ... & ... \\
PGC028887 & ...\tablefootmark{e} & ...\tablefootmark{e} & $<$1.60\tablefootmark{f} & 5.62 $\pm$ 0.15 & $<$8.03 & 7.65 & $<$7.29 & ... \\
PGC035754 & ...\tablefootmark{e} & ...\tablefootmark{e} & $<$1.39\tablefootmark{f} & $<$5.31\tablefootmark{f} & $<$7.90 & $<$7.58 & $<$7.2 & ... \\
PGC044433 & ...\tablefootmark{e} & ...\tablefootmark{e} & $<$1.63\tablefootmark{f} & $<$5.61\tablefootmark{f} & $<$7.98 & $<$7.66 & $<$7.28 & ... \\
PGC050395 & ...\tablefootmark{e} & ...\tablefootmark{e} & -0.58 $\pm$ 1.67 & 4.69 $\pm$ 0.60 & $<$7.87 & $<$7.51 & $<$7.13 & ... \\
PGC051753 & ...\tablefootmark{e} & ...\tablefootmark{e} & 0.19 $\pm$ 2.39 & 4.74 $\pm$ 0.75 & $<$7.92 & $<$7.52 & $<$7.13 & ... \\
PGC054452 & ...\tablefootmark{e} & ...\tablefootmark{e} & 0.82 $\pm$ 0.42 & 5.05 $\pm$ 0.20 & $<$7.73 & ... & ... & ... \\
PGC056772 & 98.0 $\pm$ 2.0 & 31.1 $\pm$ 0.7 & 2.45 $\pm$ 0.07 & 5.97 $\pm$ 0.07 & 8.19 $\pm$ 0.05 & ... & ... & ... \\
PGC058114 & 100.2 $\pm$ 1.8 & 25.7 $\pm$ 0.4 & 2.57 $\pm$ 0.05 & 6.48 $\pm$ 0.07 & 8.60 $\pm$ 0.02 & ... & ... & ... \\
PGC061468 & 63.5 $\pm$ 1.6 & 23.1 $\pm$ 1.1 & 2.93 $\pm$ 0.16 & 6.01 $\pm$ 0.12 & 8.00 $\pm$ 0.07 & $<$7.54 & $<$7.15 & ... \\
PGC071531 & ...\tablefootmark{e} & ...\tablefootmark{e} & $<$0.43\tablefootmark{f} & $<$5.40\tablefootmark{f} & $<$7.65 & $<$7.37 & $<$6.98 & ... \\
UGC03960 & ...\tablefootmark{e} & ...\tablefootmark{e} & 0.73 $\pm$ 0.73 & 5.31 $\pm$ 0.22 & $<$7.81 & 7.79 & 7.06 $\pm$ 0.11 & ... \\
UGC04551 & ...\tablefootmark{e} & ...\tablefootmark{e} & $<$1.47\tablefootmark{f} & 4.85 $\pm$ 0.31 & $<$7.62 & $<$7.25 & $<$6.87 & ... \\
UGC05408 & 100.3 $\pm$ 2.2 & 31.1 $\pm$ 1.5 & 2.70 $\pm$ 0.06 & 6.04 $\pm$ 0.07 & 8.32 $\pm$ 0.06 & 8.52 & 8.33 $\pm$ 0.02 & ... \\
UGC06062 & ...\tablefootmark{e} & ...\tablefootmark{e} & 1.58 $\pm$ 0.22 & $<$5.50\tablefootmark{f} & $<$7.93 & ... & ... & ... \\
UGC06176 & 82.2 $\pm$ 1.5 & 30.1 $\pm$ 0.6 & 3.34 $\pm$ 0.06 & 6.20 $\pm$ 0.08 & 8.58 $\pm$ 0.04 & 9.02 & 8.4 $\pm$ 0.02 & ... \\
UGC08876 & ...\tablefootmark{e} & ...\tablefootmark{e} & $<$1.43\tablefootmark{f} & 5.25 $\pm$ 0.21 & $<$7.80 & $<$7.43 & $<$7.05 & ... \\
UGC09519 & ...\tablefootmark{e} & ...\tablefootmark{e} & 1.50 $\pm$ 0.12 & 5.06 $\pm$ 0.09 & 8.77 $\pm$ 0.01 & 9.27 & 7.75 $\pm$ 0.02 & ... \\
\end{longtable}
\label{tab:dust}
\tablefoot{ \\
\tablefoottext{a}{Adopted from Table.~1 of \citet{you11}.} \\
\tablefoottext{b}{Adopted from Table.~B1 of \citet{ser12}.} \\
\tablefoottext{c}{Adopted from Table.~1 of \citet{you14}.} \\
\tablefoottext{d}{Adopted from Table.~3 of \citet{su15}.} \\
\tablefoottext{e}{Dust temperature not derived because the galaxy is detected in only one or no AKARI FIR band.} \\
\tablefoottext{f}{$3{\sigma}$ upper limit as the dust mass is estimated to be $0$.} \\
}
\end{landscape}
}

\end{appendix}

\end{document}